\documentclass[showpacs,twocolumn,preprintnumbers,prb,floatfix,superscriptaddress]{revtex4}

\usepackage[X2,T1]{fontenc}
\usepackage[utf8]{inputenc}
\usepackage[russian,english]{babel}
\newcommand{\ru}[1]{\foreignlanguage{russian}{#1}}

\usepackage{graphicx, color,rotating, textcomp} % Include figure files
\usepackage{dcolumn} % Align table columns on decimal point
\usepackage{bm} % bold math
\usepackage{graphicx}
\graphicspath{{pics/}} % Dort liegen die Bilder des Dokuments 
\usepackage{hyperref}
\usepackage{booktabs}

\usepackage{empheq} %mehrere aligned equations mit boxen

\usepackage{mathrsfs,amssymb} % Zusatzzeichen
\usepackage{bbm}
\usepackage{bbold}
\usepackage{amsmath}
\usepackage{amsbsy}
\usepackage{amsfonts}
\usepackage{empheq} %% allows for example to set boxes spanning multi-line equations with

\def\mathfrakG{\setbox0=\hbox{$\mathfrak{G}$}%
     \pdfliteral{1 0 .167 1 0 0 cm}\rlap{$\mathfrak{G}$}\pdfliteral{1 0 -.167 1 0 0 cm}\kern\wd0 }
\def\mathfraktG{\setbox0=\hbox{$\mathfrak{\tilde G}$}%
     \pdfliteral{1 0 .167 1 0 0 cm}\rlap{$\mathfrak{\tilde G}$}\pdfliteral{1 0 -.167 1 0 0 cm}\kern\wd0 }
\def\mathfrakF{\setbox0=\hbox{$\mathfrak{F}$}%
     \pdfliteral{1 0 .167 1 0 0 cm}\rlap{$\mathfrak{F}$}\pdfliteral{1 0 -.167 1 0 0 cm}\kern\wd0 }
\def\mathfraktF{\setbox0=\hbox{$\mathfrak{\tilde F}$}%
     \pdfliteral{1 0 .167 1 0 0 cm}\rlap{$\mathfrak{\tilde F}$}\pdfliteral{1 0 -.167 1 0 0 cm}\kern\wd0 }
\def\mathfrakA{\setbox0=\hbox{$\mathfrak{A}$}%
     \pdfliteral{1 0 .167 1 0 0 cm}\rlap{$\mathfrak{A}$}\pdfliteral{1 0 -.167 1 0 0 cm}\kern\wd0 }
\def\mathfraktA{\setbox0=\hbox{$\mathfrak{\tilde A}$}%
     \pdfliteral{1 0 .167 1 0 0 cm}\rlap{$\mathfrak{\tilde A}$}\pdfliteral{1 0 -.167 1 0 0 cm}\kern\wd0 }
\def\mathfrakB{\setbox0=\hbox{$\mathfrak{B}$}%
     \pdfliteral{1 0 .167 1 0 0 cm}\rlap{$\mathfrak{B}$}\pdfliteral{1 0 -.167 1 0 0 cm}\kern\wd0 }
\def\mathfraktB{\setbox0=\hbox{$\mathfrak{\tilde B}$}%
     \pdfliteral{1 0 .167 1 0 0 cm}\rlap{$\mathfrak{\tilde B}$}\pdfliteral{1 0 -.167 1 0 0 cm}\kern\wd0 }

\def\hG{\setbox0=\hbox{$\hat{\mathbbm{G}}\!$}%
     \pdfliteral{0.8 0 .1336 1 0 0 cm}\rlap{$\hat{\mathbbm{G}}$}\pdfliteral{1.25 0 -.167 1 0 0 cm}\kern\wd0 }
\def\cG{\setbox0=\hbox{$\check{\mathbbm{G}}\!$}%
     \pdfliteral{0.8 0 .1336 1 0 0 cm}\rlap{$\check{\mathbbm{G}}$}\pdfliteral{1.25 0 -.167 1 0 0 cm}\kern\wd0 }

\newcommand{\DDelta}{\mathit{\Delta}}

\renewcommand{\b}[1]{\boldsymbol{#1}}

\begin{document}

% Title Page
\title{Theory of a Weak-Link Superconductor-Ferromagnet Josephson Structure}

\author{J. Gelhausen}
\affiliation{Institut f\"ur Theoretische Physik, Universit\"at zu K\"oln, 50937 Cologne, Germany}
\affiliation{Department of Physics, Royal Holloway, University of London, Egham, Surrey TW20 0EX, United Kingdom}
\author{M.~Eschrig}
\affiliation{Department of Physics, Royal Holloway, University of London, Egham, Surrey TW20 0EX, United Kingdom}
\date{\today}

\begin{abstract}
\noindent
We propose a model for the theoretical description of a weak-link Josephson junction, in which the weak link is spin-polarized due to proximity to a ferromagnetic metal (S-(F$|$S)-S).
Employing Usadel transport theory appropriate for diffusive systems, we show that the weak link is described within the framework of Andreev circuit theory by an effective self-energy resulting from the implementation of spin-dependent boundary conditions. This leads to a considerable simplification of the model, and allows for an efficient numerical treatment.
As an application of our model, we show numerical calculations of important physical observables such as the local density of states, proximity-induced minigaps, spin-magnetization, and the phase and temperature-dependence of Josephson currents of the S-(F$|$S)-S system. We discuss 
multi-valued current-phase relationships at low temperatures as well as their crossover to sinusoidal form at high temperatures. Additionally, we numerically treat (S-F-S) systems that exhibit a magnetic domain wall in  the F region and calculate the temperature-dependence of the critical currents.
\end{abstract}
\maketitle

\newcommand{\pF}{\vec{p}_{\rm F}}
\newcommand{\vphi}{\varphi}
\newcommand{\eps}{\varepsilon}
\newcommand{\ud}{\uparrow,\downarrow}
\renewcommand{\u}{\uparrow}
\renewcommand{\d}{\downarrow}
\newcommand{\ket}[1]{| {#1}\rangle}
\newcommand{\bra}[1]{\langle {#1}|}
\newcommand{\barlambda}{{\lambda \!\!\!^{-}\,\!}}
\newcommand{\blFe}{{\lambda \!\!\!^{-}\,\!}_{\mathrm{F}1}}
\newcommand{\blFeta}{{\lambda \!\!\!^{-}\,\!}_{\mathrm{F}\eta}}
\newcommand{\blF}{{\lambda \!\!\!^{-}\,\!}_{\mathrm{F}}}
\newcommand{\blJ}{{\lambda \!\!\!^{-}\,\!}_{J}}
\newcommand{\EF}{E_{\mathrm{F}}}
\newcommand{\vpfe}{\vec{p}_{\mathrm{F}1}}
\newcommand{\vpfz}{\vec{p}_{\mathrm{F}2}}
\newcommand{\vpfd}{\vec{p}_{\mathrm{F}3}}
\newcommand{\vpfeta}{\vec{p}_{\mathrm{F}\eta }}
\newcommand{\vvfe}{\vec{v}_{\mathrm{F}1}}
\newcommand{\vvfz}{\vec{v}_{\mathrm{F}2}}
\newcommand{\vvfd}{\vec{v}_{\mathrm{F}3}}
\newcommand{\vvfzd}{\vec{v}_{\mathrm{F}2,3}}
\newcommand{\vvfeta}{\vec{v}_{\mathrm{F}\eta }}
\newcommand{\pfe}{p_{\mathrm{F}1}}
\newcommand{\pfz}{p_{\mathrm{F}2}}
\newcommand{\pfd}{p_{\mathrm{F}3}}
\newcommand{\pfeta}{p_{\mathrm{F}\eta }}
\newcommand{\vfe}{v_{\mathrm{F}1}}
\newcommand{\Nfe}{N_{\mathrm{F}1}}
\newcommand{\vfz}{v_{\mathrm{F}2}}
\newcommand{\Nfz}{N_{\mathrm{F}2}}
\newcommand{\vfd}{v_{\mathrm{F}3}}
\newcommand{\Nfd}{N_{\mathrm{F}3}}
\newcommand{\vfzd}{v_{\mathrm{F}2,3}}
\newcommand{\vfeta}{v_{\mathrm{F}\eta}}
\newcommand{\JFM}{J_{\mathrm{FM}}}
\newcommand{\vJFM}{\vec{J}_{\mathrm{FM}}}
\newcommand{\JI}{J_{\mathrm{I}}}
\newcommand{\vJI}{\vec{J}_{\mathrm{I}}}
\newcommand{\VI}{V_{\mathrm{I}}}
\newcommand{\gr}{\gamma^R}
\newcommand{\ga}{\gamma^A}
\newcommand{\grt}{\tilde{\gamma}^R}
\newcommand{\gat}{\tilde{\gamma}^A}
\newcommand{\gra}{\gamma^{R,A}}
\newcommand{\grat}{\tilde{\gamma}^{R,A}}
\newcommand{\Rs}{{R}_1}
\newcommand{\Tsn}{{T}_{sn}}
\newcommand{\Tns}{{T}_{ns}}
\newcommand{\Rn}{{R}_n}
\newcommand{\ruu}{r_{2}}
\newcommand{\rdd}{r_{3}}
\newcommand{\Tsu}{{T}_{12}}
\newcommand{\Tsd}{{T}_{13}}
\newcommand{\Tus}{{T}_{21}}
\newcommand{\Tds}{{T}_{31}}
\newcommand{\rud}{r_{23}}
\newcommand{\rdu}{r_{32}}
\newcommand{\vecb}[1]{\mathbf{#1}}
\newcommand{\Tc}{T_\mathrm{c}}
\newcommand{\h}[1]{\hat{#1}}
\renewcommand{\l}{\left(}
\renewcommand{\r}{\right)}
\newcommand{\lr}[1]{\left( #1 \right)}
\newcommand{\Fig}[1]{Fig.\!\! \ref{#1}}
\newcommand{\Eq}[1]{Eq.\!\! \!(\ref{#1})}

\section{Introduction}

The study of superconductivity in proximity with ferromagnetic materials has opened the path towards creation and control of spin-polarized Cooper pairs and superconducting spin currents.\cite{Izyumov02,Eschrig04,Golubov04,Bergeret05a,Buzdin05,Lyuksyutov07,Eschrig11,Robinson14,Eschrig15,Linder15} 
Recent developments show that also energy currents can be managed 
%thermoelectric effects can be enhanced drastically 
by using spin-polarized Cooper pairs \cite{Machon13,Kawabata13,Machon14,Ozaeta14,Giazotto15}.
A considerable amount of work has concentrated on spin-polarized supercurrents across ferromagnetic metals or insulators. 
Hybrid structures in which superconductors are connected by a weak link of a normal-metal/ferromagnet bilayer or a ferromagnet/normal-metal/ferromagnet trilayer forming a bridge between the superconducting banks have been studied to a lesser extend.
Theoretical proposals for such structures\cite{Karminskaya07} have been followed by experimental work on hybrid planar Al-(Cu$|$Fe)-Al submicron bridges\cite{Golikova12}, and by further theoretical investigations to optimize practical performance.\cite{Karminskaya10}

\begin{figure}[b]
\includegraphics[width=0.9\columnwidth]{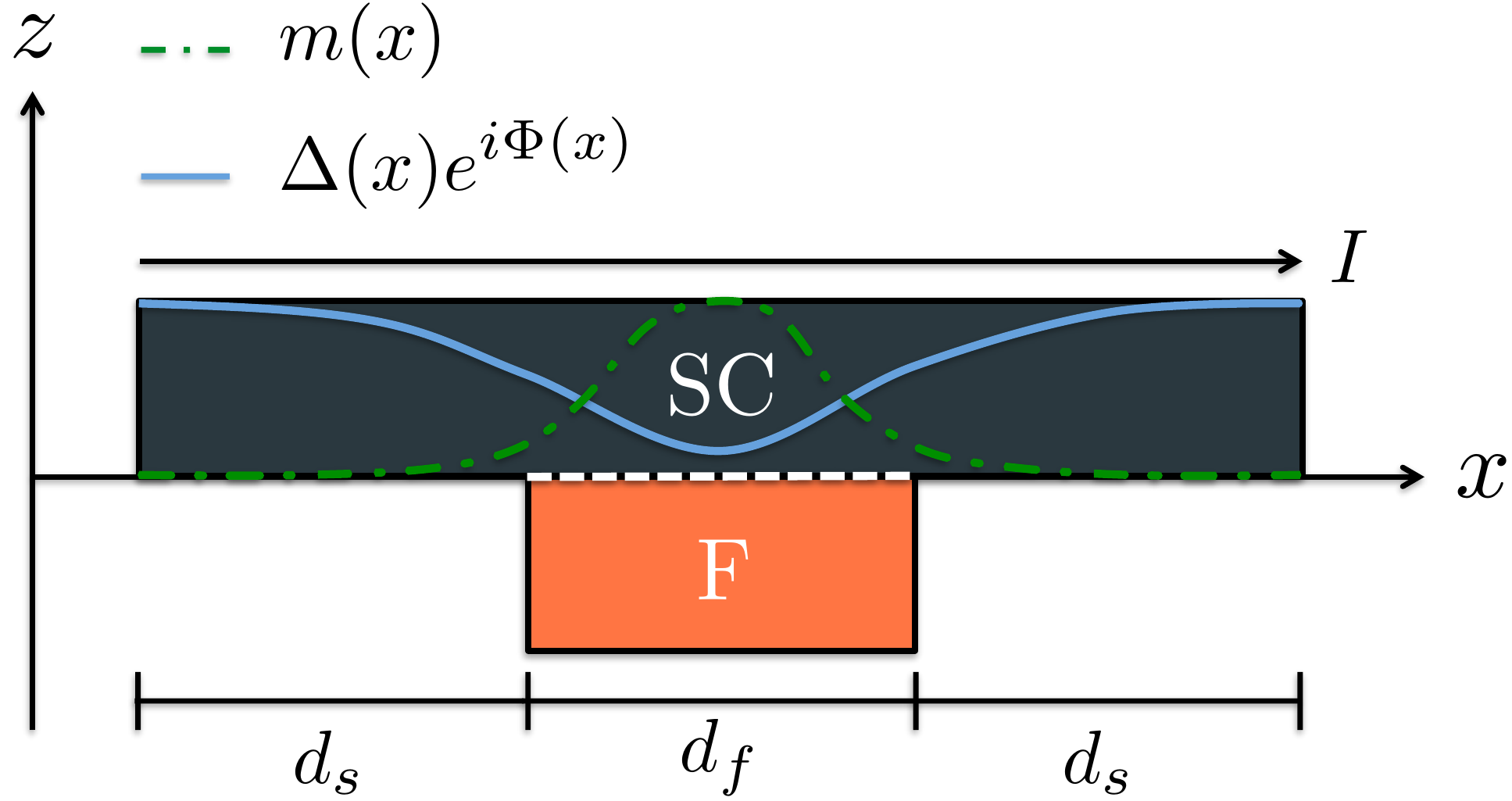}
\caption{
\label{setup} 
Model of an S-(S|F)-S junction. The dashed line between the superconductor (SC) and the ferromagnet (F) indicates a spin-active interface. The structure is of length $2d_s+d_f$, where $d_s$ is the length of the superconducting banks on either side of the ferromagnet, and
$d_f$ is the length of the superconductor/ferromagnet proximity block. The pair potential $\Delta e^{i\Phi}$ (blue, solid) and the spin-magnetization $m$ (green, dot-dashed) are shown schematically as a function of spatial coordinate $x$. A supercurrent $I$ is driven by a spatially varying superconducting phase $\Phi(x)$.}
\label{FIG:1}
\end{figure}

In the present work we study the case of a weak link consisting of 
a superconductor/ferromagnetic-metal bilayer, where the 
superconducting material is the same as in the leads, and where superconductivity is suppressed due to proximity coupling to the ferromagnetic metal, e.g. as in an Al-(Al$|$Co)-Al structure.
A schematic illustration of the system is depicted in \Fig{FIG:1}; 
an experimental realization could be a superconducting strip running across a ferromagnetic disc.
In our modeling, the structure consists in total of two blocks; the superconductor and the ferromagnet which are connected by an interface over a length $d_f$ (dashed line in Fig. \ref{setup}), building the weak link. 
A superconductor in proximity with a ferromagnet exhibits spin-polarized Cooper pairs, which can be considered as a mixture between spin-singlet and spin-triplet pairs. 
This in turn implies a spin-polarized excitation spectrum,
resulting in a spin-magnetization of the superconductor in the region where it is proximity-coupled to the ferromagnet\cite{Alexander85,Tokuyasu88,Bergeret04}
(see dashed-dotted line in Fig. \ref{setup}).
The singlet superconducting order parameter is shown in Fig. \ref{setup} as full line, exhibiting the suppression in the proximity-coupled region. 

We employ a Green function technique for metals, itinerant ferromagnets, and superconductors in the diffusive limit. 
Within this theory, Green functions are described by transport equations of the kind derived by Usadel,\cite{Usadel} generalized to spin-dependent phenomena within a Riccati representation.\cite{Eschrig04,Konstandin05,Cuevas06}
The considered structure makes it necessary to self-consistently calculate the pair potential with the spectrum of excitations (as encoded by the Green functions). 
Due to the presence of the ferromagnet, we supplement the transport equations for the Green functions with spin-dependent boundary conditions.\cite{Eschrig15a}
We propose a model in which spin-dependent interface scattering phase shifts \cite{Tokuyasu88,Cottet09}
lead to a spin-polarization of Cooper pairs in the superconducting regions of the weak link. Assuming the thickness of the superconductor within the weak link much smaller than the superconducting coherence length,
we are able to cast the boundary conditions in the form of an effective self-energy, which enters a one-dimensional transport equation in direction of the weak link.

In Section \ref{Sec:TransportSSFSSystem} we present the theoretical framework to describe our model. We supplement the Usadel equation by a self-energy-like contribution that is derived in the framework of an Andreev circuit theory to account for the spin-dependent boundary conditions.

In Section \ref{Sec:Observables} we calculate characteristic observables such as the local density of states, the spin magnetization of the system, the superconducting order parameter, the characteristic current-phase relationship, and the temperature-dependence of the critical Josephson current. All calculations are performed self-consistently. 

In Section \ref{Sec:CurrCons} we explicitly show that our model fulfills the requirement of charge conservation.

In Section \ref{Sec:SFSMagneticDomainWall} we numerically investigate an S-F-S heterostructure that exhibits a magnetic domain wall. We extend previous work\cite{Konstandin05} 
by a self-consistent calculation of the pair-potential, and calculate the local density of states, current-phase relations, and the temperature-dependent critical current.

\section{Theoretical Description}
\label{Sec:TransportSSFSSystem}
We employ for our theoretical treatment Usadel theory of diffusive superconductors,\cite{Usadel,Belzig99} adapted for spin-polarized systems (see e.g. Ref. \onlinecite{Eschrig15}).
Usadel theory can be derived from the theory of Eilenberger \cite{Eilenberger68} and of Larkin and Ovchinnikov \cite{Larkin68} in the diffusive limit.
The equilibrium physics is captured by the
retarded Green function (or propagator) $\hG^{R} \equiv \hG({\bf R},\epsilon)$ where ${\bf R}$ denotes the spatial coordinate, ${\bf R}=(x,y,z)$, and $\epsilon$ the energy. Current transport will be considered in $x$-direction, whereas $z$ denotes the direction perpendicular to the superconducting films.
The propagator
$\hG({\bf R},\epsilon)$ has a total of $16$ complex-valued components and is build up of four $2 \times 2$ block spin-matrices, two of which are related to the other two by particle-hole conjugation symmetry.
This matrix structure arises from the internal degrees of freedom: the spin degree of freedom and the particle-hole degree of freedom. 
The hat accent denotes the 2$\times$2 block matrix structure in particle-hole (Nambu-Gor'kov) space: 
\begin{align}
\hG=
\left(
\begin{array}{cc}
\mathfrakG & \mathfrakF \\
 \mathfraktF & \mathfraktG \\
\end{array}
\right)
\label{Eq:RetardedGreenFunction}
\end{align}
where $\mathfrakG $, $\mathfrakF$, $\mathfraktG$, and $\mathfraktF$ are 2$\times$2 spin matrices, i.e.
$\mathfrakF_{\alpha \beta}$ has spin indices $\alpha, \beta = \{\uparrow, \downarrow\}$ etc.
The off-diagonal elements $\mathfrakF$ and $\mathfraktF$ quantify the superconducting pair correlations.
The propagator can be analytically continued from the real energy axis into the upper complex half plane, $\epsilon \to \varepsilon $ with Im$(\varepsilon )\ge 0$.
The symmetry relation (particle-hole conjugation) between the block spin-matrices is given by the ``tilde''-operation:
\begin{align}
 \mathfraktA({\bf R},\varepsilon)=\mathfrakA ({\bf R},-\varepsilon^{*})^{*}
 \label{EqTildeOperation}
 \end{align}
where $(^{*})$ denotes complex conjugation.

In addition to the discreet internal degrees of freedom, there are continuous external degrees of freedom, which are described by the energy $\epsilon $ and the spatial coordinate ${\bf R}$.
The diffusive motion is described by a quantum kinetic transport equation, in our case the Usadel equation, \cite{Usadel} which for the propagator 
within the superconductor, $\hG_{Sc}({\bf R},\epsilon)$,
takes the form
\begin{align}
\left[\epsilon \h{\tau}_3-\h{\Delta} ,\hG_{Sc}({\bf R},\epsilon)\right]+\frac{D}{\pi}\nabla \left(\hG_{Sc}({\bf R},\epsilon)\nabla \hG_{Sc}({\bf R},\epsilon)\right)=\hat 0
\label{EQ:UsadelEquation}
\end{align}
where $[\h A,\h B]\equiv \h A\h B-\h B\h A$,
the 4$\times $4 matrix $\hat\tau_3$ is the direct product between the
third Pauli matrix in particle-hole space and the spin unit matrix, $\h 0$ is the 4$\times $4 zero matrix,
$\nabla \equiv \partial/\partial {\bf R}$, and $D$ is the diffusion constant.
This transport equation is supplemented by the normalization condition
\begin{align}
\hG_{Sc}({\bf R},\epsilon)^2=-\pi^2 \h 1,
\end{align}
where $\h 1\equiv \mathbb{1}_{4x4}$ is the 4$\times$4 unit matrix.

For the system depicted in \Fig{setup} we make a simplifying ansatz that allows us to transform the Usadel equation into a quasi-one dimensional differential equation,
supplemented by a self-energy-like contribution that accounts for the influence of the ferromagnet on the superconductor.
This ansatz is motivated by assuming that the superconductor of thickness $d$ does not extend significantly in the $z$ direction, meaning that the spatial variations of the superconducting order parameter in the $z$-direction are small. This is justified for example for a superconducting strip whose lateral dimensions are much bigger than its vertical extension. Our perturbative ansatz for the Green function of the system depicted in \Fig{FIG:1} is thus:
\begin{align}
\hG_{Sc}(x,z,\epsilon)=\hG_{0}(x,\epsilon)+\hG_1(x,\epsilon)(z-d)^2
\label{EqAnsatzforGreenfunction}
\end{align}
with the normalization condition
\begin{align}
\hG_{Sc}(x,z,\epsilon)^2&=\l\hG_{0}(x,\epsilon)+\hG_1(x,\epsilon)(z-d)^2\r^2 \nonumber \\
&=-\pi^2 \h 1 \quad \forall (x,z,\epsilon).
\end{align}
Up to linear order in $(z-d)$ this means
\begin{align}
\hG_{0}(x,\epsilon)^2&=-\pi^2 \h 1 \quad \forall (x,\epsilon) .
\label{EQ:Green-FunctionNormalisation}
\end{align}
The surfaces at $z=d$ border to an insulating $(I)$ region. The boundary conditions at the $S/I$ interface must satisfies Nazarov's boundary conditions \cite{NazarovYuli} $\partial_{z} \hG_{Sc}(x,z=d,\epsilon)=0$. A linear contribution of the form $(z-d)$ in Eq. \eqref{EqAnsatzforGreenfunction} does not satisfy this condition and therefore the ansatz for the spatial variation in the $z$-direction contains only a quadratic contribution, proportional to $(z-d)^2$.

To leading order in $(z-d)$ the Usadel equation reads
\begin{align}
[\epsilon \h{\tau}_3-\h{\Delta } ,\hG_{0}(x,\epsilon)]&+\frac{D}{\pi}\partial_x[\hG_{0}(x,\epsilon)\partial_x\hG_{0}(x,\epsilon)]\nonumber \\
&+2\frac{D}{\pi}\hG_{0}(x,\epsilon)\hG_{1}(x,\epsilon)=\hat 0.
\label{EqUsadelwithContribution}
\end{align}
The contribution $\hG_1(x,\epsilon)$ will be determined from the boundary conditions of the problem and will thus depend on the structure of the ferromagnet. A detailed derivation of this expression can be seen further below, in Eq. \eqref{EQ:G1Contribution}.
Here we note that we will show that 
the Usadel Equation \eqref{EqUsadelwithContribution} can be cast into the form
\begin{align}
&[\epsilon \h{\tau}_3-\h{\Delta} -\h{\Sigma}(x,\epsilon),\hG_{0}(x,\epsilon)]
\nonumber \\ &
\qquad \qquad+\frac{D}{\pi}\partial_x[\hG_{0}(x,\epsilon)\partial_x\hG_{0}(x,\epsilon)]=\hat{0},
\label{EqNewUsadelEquationWithSigma}
\end{align}
where $\h{\Sigma}(x,\epsilon)$ formally appears like a self-energy contribution to the system and captures the influence of the ferromagnet. It is defined in Eq. \eqref{EQ:DefinitionofSigma} below.

\subsection{Riccati Parameterization}
The Green functions can be described in the framework of the spin-dependent Riccati parameterization.\cite{Eschrig00} This parameterization allows to retain the full spin structure of the Green function while automatically ensuring the normalization condition. The power of this parameterization for diffusive systems was exemplified, for example, by calculating the effects of the superconducting proximity effect through magnetic domain walls.\cite{Konstandin05}   Within this framework the retarded Green function $\hG_0(x,\epsilon)$ is parameterized by
\begin{equation}
\hG_0=-i\pi\hat{N} 
\left(
\begin{array}{cc}
 (\mathit{1}+\gamma  \tilde{\gamma}) & 2\gamma\\
-2\tilde{\gamma} & -(\mathit{1}+\tilde{\gamma} \gamma) \\
\end{array}
\right) 
\label{EqGreenFunction}
\end{equation}
where $\mathit{1}$ is the spin unit matrix, and where
\begin{equation}
\hat{N}=\left(
\begin{array}{cc}
 (\mathit{1}-\gamma  \tilde{\gamma})^{-1} & 0\\
0 & (\mathit{1}-\tilde{\gamma} \gamma)^{-1} \\
\end{array}
\right) 
\label{EQ:Riccatti2}
\end{equation}
automatically ensures the normalization condition \eqref{EQ:Green-FunctionNormalisation}. The coherence functions $\gamma$ and $\tilde{\gamma}$ are spin matrices, $\gamma_{\alpha\beta}$ with $\alpha,\beta=\{\uparrow,\downarrow\}$, where each element depends on the energy $\epsilon$ and the spatial coordinate $x$. 

We now write the transport equations Eq.\! \eqref{EqNewUsadelEquationWithSigma} in the Riccati parameterization.
The $4 \times 4$ matrix $\hat\Sigma(x,\epsilon)$ is only non-zero in the range where the proximity effect between the superconductor and the ferromagnet is in action,
and can be written in $2 \times 2$ block structure 
\begin{align}
\h{\Sigma}(x,\epsilon)&=\left(
\begin{array}{cc}
 \mathfrakA & \mathfrakB \\
 \mathfraktB & \mathfraktA \\
\end{array}
\right).
\end{align}
With this definition, the Usadel equations for the coherence functions $\gamma$ and $\tilde{\gamma}$ are written as \cite{Eschrig04,Konstandin05}
\begin{align}
&\frac{d^2\gamma}{dx^2}+\left(\frac{d\gamma}{dx}\right) \frac{\mathfraktF }{i\pi}  \left(\frac{d\gamma}{dx}\right)= \nonumber \\
&\; \frac{i}{D}\left[\gamma   ( \DDelta^{*}+\mathfraktB )  \gamma-\l \epsilon\mathit{1}-\mathfrakA  \r   \gamma-\gamma  (\epsilon \mathit{1}+\mathfraktA )-\DDelta-\mathfrakB \right], \label{EQ:Usadelgamma}\\
&\frac{d^2\tilde{\gamma}}{dx^2}+\left(\frac{d\tilde{\gamma}}{dx}\right)   \frac{\mathfrakF }{-i\pi}  \left(\frac{d\tilde{\gamma}}{dx}\right)= \nonumber \\
&\; \frac{-i}{D}\left[\tilde{\gamma}   \l\DDelta+\mathfrakB \r  \tilde{\gamma}+(\epsilon\mathit{1}+\mathfraktA )  \tilde{\gamma}+\tilde{\gamma}  (\epsilon \mathit{1}-\mathfrakA )-\DDelta^{*}-\mathfraktB \right]
\label{EQ:Usadelgamma2}
\end{align}
with
\begin{align}
\hat{\tau}_3=
\left(
\begin{array}{cc}
 \mathit{1} & 0\\
 0 & -\mathit{1} \\
\end{array}
\right),
 \quad \hat{\sigma}_i=
\left(
\begin{array}{cc}
 \sigma_i & 0\\
 0 & \sigma^{*}_i \\
\end{array}
\right),
\quad \hat{\Delta}=
\left(
\begin{array}{cc}
 0 & \DDelta\\
 \DDelta^{*} & 0 \\
\end{array}
\right) 
\end{align}
where
$\sigma_i$ are the Pauli spin-matrices with $i=x,y,z$. The (temperature-dependent) spin-singlet superconducting order-parameter is given by
\begin{align}
\DDelta(x)=\Delta(x) i \sigma_y=
\Delta_0(x)
e^{i\Phi(x)}
\cdot 
i\sigma_y
\end{align}
where $\Delta_0(x)$ is the modulus of the order parameter, and
$\Phi(x)$ denotes a spatially dependent, real phase. The order parameter 
$\Delta(x)=\Delta_0(x) e^{i \Phi(x)}$
must be determined self-consistently as described further below, to ensure current conservation across the weak link.
The Usadel equation must be supplemented by appropriate
boundary conditions. This will be addressed in the next section.

\subsection{Andreev circuit theory}
\label{Sec:Spin-dependentBoundaryCondition}

We wish to employ spin-dependent boundary conditions to couple the ferromagnet to the superconductor.
A crucial quantity at a boundary between a strongly spin-polarized ferromagnet and a superconductor is the spin-mixing parameter \cite{Tokuyasu88}, or spin-mixing conductance ${\cal G}_\phi $\cite{Brataas00,Cottet09,Machon13,Eschrig15a}. This parameter is the crucial quantity leading to spin-polarization of Cooper pairs as well as to a spin-split local density of states at the contact, 
and results from spin-dependent scattering phase shifts during reflection and transmission at a superconductor-ferromagnet interface.\cite{Tokuyasu88,Brataas00,Cottet09}

\begin{figure}[t]
\includegraphics[width=6cm]{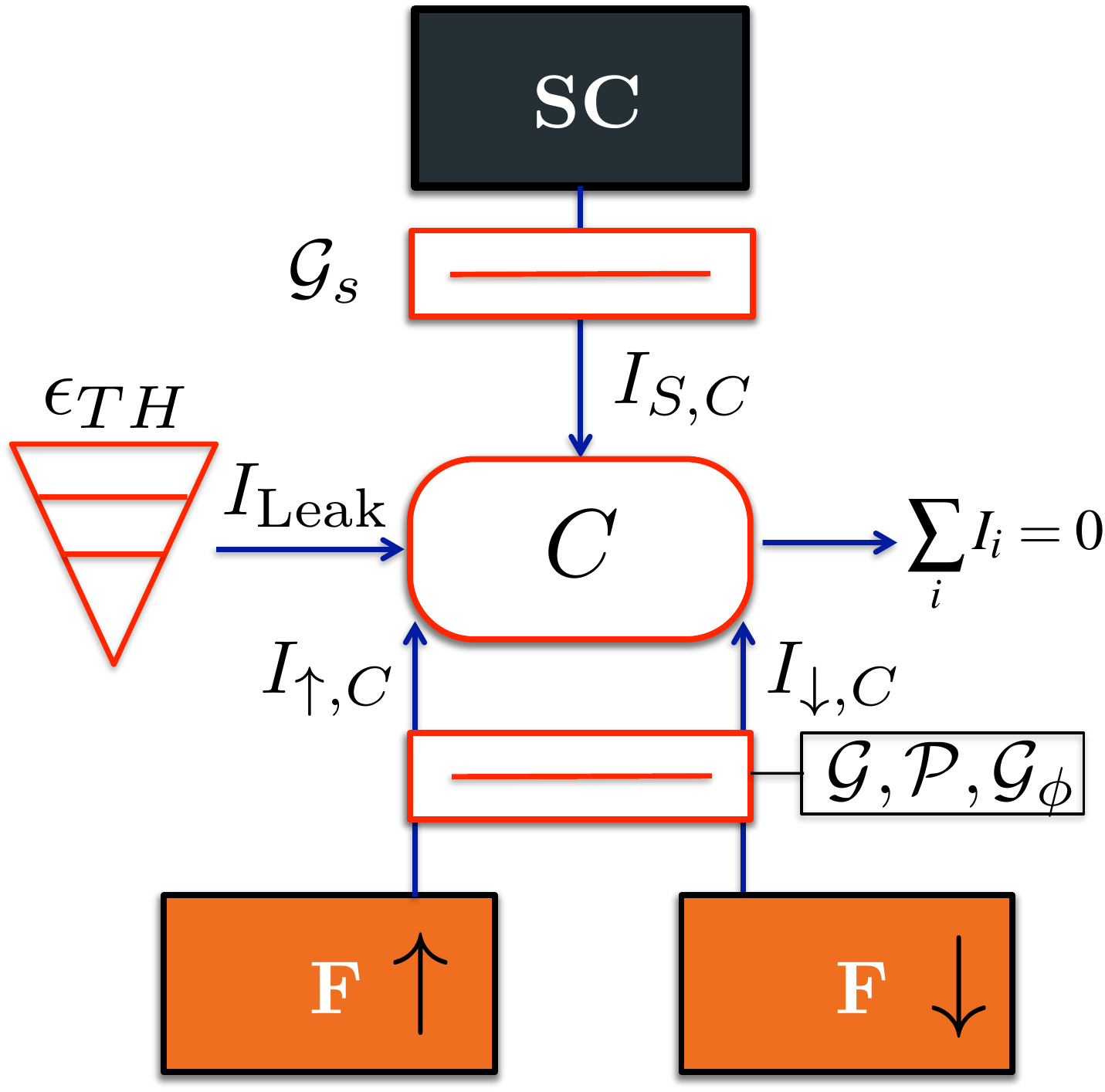}
\caption{\label{FIG:2} Illustration of an Andreev circuit to calculate the Green function $\hG_{C}$ of the central node $C$. It consists of a central node $C$ that is connected to a ferromagnetic terminal by a connector that is characterized by a set of conductance parameters ${\cal G} $, ${\cal P}$, ${\cal G}_{\phi}$ (see text). The leakage terminal is characterized by a Thouless energy $\epsilon_{\rm Th}$. The arrows between the blocks indicate the flow of matrix currents that obey a Kirchhoff rule, see Eq.\! \eqref{EqKirchhoffRule}.}
\end{figure}

\begin{figure}[b]
\includegraphics[width=6cm]{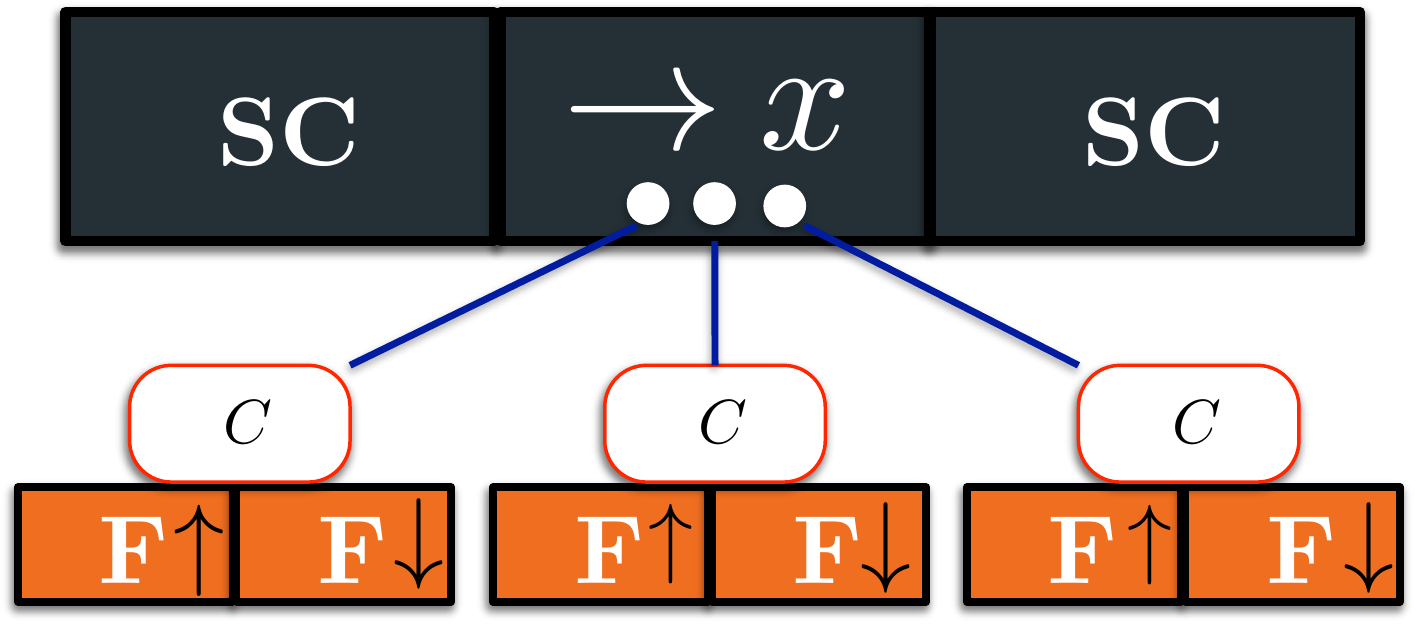}
\caption{\label{FIG:3} The Green function $\hG_C$ of the central node has to be calculated for every point $x$ at the interface and for all energies. The red blocks labeled $C$ and $F$ refer to the central node $C$ and the ferromagnet blocks discussed in \Fig{FIG:2}.}
\end{figure}

In order to implement boundary conditions, we utilize a discretized (Andreev) quantum circuit theory \cite{NazarovYuli,NazarovYuli2} where the system consists of terminals, nodes, and connectors, as depicted in Figs. 
\ref{FIG:2} and \ref{FIG:3}.
%\ref{FIG:3} and \ref{FIG:2}.
Within the proximity region, at each spatial point $x$ the superconductor is tunnel-coupled to a central node $C$. This coupling is characterized by a boundary conductance ${\cal G}_S$. The node itself is in contact to a ferromagnetic metal via a spin-dependent coupling that is characterized by its polarization ${\cal P}$, its boundary conductance value ${\cal G}$, and the spin-mixing parameter ${\cal G}_{\phi}$.

The loss of superconducting correlations is accounted for by a leakage current that contains the Thouless energy $\epsilon_{\rm Th}$ of the leakage terminal. The central node $C$ is responsible to model the behavior of the superconducting correlations in the structure under the effect of leakages and spin-polarized boundaries of the ferromagnet. 
As has been shown by Nazarov \cite{NazarovYuli}, the following generalized Kirchhoff rule for the
so-called matrix current holds (see Fig. \ref{FIG:2}):
\begin{equation}
\hat{I}_{S,C}+\hat{I}_{\downarrow,C}+\hat{I}_{\uparrow,C}+\hat{I}_{\rm Leak}=\h 0
\label{EqKirchhoffRule}
\end{equation}
where $\hat{I}_{S,C}$ is the matrix current from the superconductor to the node $C$. The matrix currents from the ferromagnet into the central node $C$ are denoted by $\hat{I}_{\downarrow,C}, \hat{I}_{\uparrow,C}$ whereas $\hat{I}_{\rm Leak}$ is the matrix current from the leakage terminal going into the central node. Eq.\! \eqref{EqKirchhoffRule} has to be applied at each interface point $(x,z)$ with $z=0$ at the interface between the superconductor and the ferromagnet (see Fig.~\ref{FIG:3}).

The leakage current is given by
\begin{align}
\hat{I}_{\rm Leak}(x,\epsilon)=&\frac{{\cal G}_q}{4 \epsilon_{\rm Th}}[\hG_{\text{Leak}}(\epsilon),\hG_C(x,\epsilon)]\label{EQ:LeakTerminal}
\end{align}
where $\hG_{\text{Leak}}(\epsilon)=-\pi \epsilon \h{\tau}_3$
is an energy-dependent quantity to account for a leakage of coherence. All information about the leakage terminal is given by its Thouless energy $\epsilon_{\text{Th}}$. 

The matrix current between the terminals $j$ and the central node $C$ in linear order in $\mathcal{T}_{n}$ (see below) can be written in the form of the following commutator\cite{Machon13,Machon14}
\begin{align}
\hat{I}_{j,C}=&\frac{1}{2}\bigg[{\cal G}_{0,j} \cdot \hG_{j}+{\cal G}_{P,j}\cdot \{\hat{\kappa}_j,\hG_j\} \bigg.\nonumber \\& \qquad \bigg.
+{\cal G}_{1,j} \cdot \hat{\kappa}_j \hG_j\hat{\kappa}_j-\pi {\cal G}_{\phi,j}\cdot \hat{\kappa}_j,\hG_{C}\bigg]
\label{EQ:GeneralMCurrent}
\end{align}
where $j \in\{\uparrow, \downarrow, S\}$ labels the terminal.
The boundary conditions are specified by the set of conductance parameters\cite{Machon13,Machon14}
\begin{align}
{\cal G}_{0}&={\cal G}_q\sum\limits_{n}^{\# \text{Channels}}\mathcal{T}_{n}\l 1+\sqrt{1-\mathcal{P}_n^{2}} \r \label{EQGGPDELTAGPHI1}\\
{\cal G}_{1}&={\cal G}_q\sum\limits_{n}^{\# \text{Channels}}\mathcal{T}_{n}\l 1-\sqrt{1-\mathcal{P}_n^{2}} \r\\
{\cal G}_{P}&={\cal G}_q\sum\limits_{n}^{\# \text{Channels}}\mathcal{T}_{n}\mathcal{P}_{n}\\
{\cal G}_{\phi}&=2{\cal G}_q\sum\limits_{n}^{\# \text{Channels}} \delta \Phi_n .
\label{EQGGPDELTAGPHI}
\end{align}
Here, the spin-mixing parameter is described by $\delta \Phi_n$, the spin polarization of the ferromagnet by $\mathcal{P}_{n}$, the spin-averaged transmission probability for channel $n$ is given by $\mathcal{T}_n$ and ${\cal G}_q=\frac{e^2}{h}$ is the quantum conductance.

For a strongly spin-polarized ferromagnet the transmission and reflection channels at the interface are completely spin-polarized (${\cal P}_n=1$ and $-1$ for spin-up and spin-down, respectively), such that we obtain
\begin{align}
{\cal G}_{0}&={\cal G}_{1}=
{\cal G}_q\left(\sum\limits_{n}^{\#\uparrow}\mathcal{T}_{n}
+\sum\limits_{n}^{\#\downarrow}\mathcal{T}_{n} \right)
\equiv \frac{1}{2} ({\cal G}_\uparrow+{\cal G}_\downarrow )
\\
{\cal G}_{P}&=
{\cal G}_q\left(\sum\limits_{n}^{\#\uparrow}\mathcal{T}_{n}
-\sum\limits_{n}^{\#\downarrow}\mathcal{T}_{n} \right)
\equiv \frac{1}{2} ({\cal G}_\uparrow-{\cal G}_\downarrow )
\\
{\cal G}_{\phi}&=2{\cal G}_q
\left(\sum\limits_{n}^{\#\uparrow}\delta \Phi_{n}
-\sum\limits_{n}^{\#\downarrow}\delta \Phi_{n} \right)\equiv {\cal G}_{\phi,\uparrow}-{\cal G}_{\phi,\downarrow }
\end{align}
where $\#\uparrow $ is the number of spin-up channels and $\#\downarrow $ the number of spin-down channels.

We obtain the matrix current between the superconductor and the central node $C$ 
by setting ${\cal G}_{1,S}={\cal G}_{P,S}={\cal G}_{\phi,S}=0$ and defining ${\cal G}_{0,S}={\cal G}_{S}$ for the superconductor. 
The expression for the ferromagnetic contacts is simplified by $[\hG_j,\h{\kappa}_j]=0$; thus, one can combine ${\cal G}_{\uparrow,(\downarrow)}={\cal G}_{0,\uparrow(\downarrow)}+{\cal G}_{1,\uparrow(\downarrow)}$. 
Furthermore, 
we simplify the notation by setting $\h{I}_{j,C}\equiv \h{I}_j$. 
The various matrix currents are then determined by the following boundary conditions \cite{Machon13,Eschrig15}:
\begin{align}
\h{I}_{\alpha}(x,\epsilon)=&\frac{1}{2}\big[{\cal G}_{\alpha } \cdot \hG^{F}_{\alpha}+{\cal G}_{P,\alpha}\cdot \{\h{\kappa}_{\alpha}, \hG^{F}_{\alpha}\} \big.
\nonumber\\ &
\qquad \qquad \big. -\pi {\cal G}_{\phi,\alpha }\cdot \h{\kappa}_{\alpha},\hG_C(x,\epsilon)\big]\label{EQ:alphaTerminal}\\
\h{I}_S(x,\epsilon)=&\frac{1}{2}{\cal G}_S \cdot [\hG_{0}(x,\epsilon),\hG_C(x,\epsilon)]\label{EQ:ScTerminal}
\end{align}
where $\alpha \in \left\{\uparrow,\downarrow\right\}$, and
\begin{align}
\h{\kappa}_{\uparrow}&=-\h{\kappa}_{\downarrow}\equiv\h{\kappa}=\mathbbm{1}_{2\times 2} \otimes ({\bf m} \cdot {\b \sigma}),\\
\hG^{F}_{\uparrow}&=\hG^{F}_{\downarrow}=-i\pi \h{\tau}_3\equiv \hG^F.
\end{align}
Here, $\hG^F$ is the solution to the Usadel equation \eqref{EQ:UsadelEquation} for a non-superconducting material ($\hat{\Delta}=\hat{0}$).
The direction of the magnetization of the ferromagnet is described by the spin-matrix $\h{\kappa}$, where $\b{m}$ is the unit vector of magnetization of the interface and $\b{\sigma}$ is the vector of spin Pauli matrices. $\hat\tau_0$ is the unit matrix in 2$\times$2 Nambu-Gor'kov space. The Green function $\hG_0(x,\epsilon)$ is the Green function defined in Eq.\! \eqref{EqGreenFunction} that solves Eq.\! \eqref{EQ:Usadelgamma} and Eq.\! \eqref{EQ:Usadelgamma2}.

More compactly, we can write:
\begin{align}
&\h{I}_F(x,\epsilon)=\hat{I}_{\downarrow}+\hat{I}_{\uparrow}\nonumber \\
&\qquad =\frac{1}{2}[{\cal G} \hG^{F}+{\cal G}_{P}\{\h{\kappa}, \hG^{F}\}-\pi {\cal G}_{\phi}\h{\kappa},\hG_C(x,\epsilon)],
\end{align}
and the boundary of the ferromagnet to the superconductor is characterized by the three parameters ${\cal P}$, ${\cal G}$, and ${\cal G}_{\phi}$, see \Fig{FIG:2},
\begin{align}
\label{C1}
{\cal G}&={\cal G}_{\uparrow}+{\cal G}_{\downarrow},\\
{\cal G}_{P}&={\cal G}_{P,\uparrow}+ {\cal G}_{P,\downarrow}=\frac{1}{2}\l {\cal G}_{\uparrow}-{\cal G}_{\downarrow}\r=\frac{1}{2}{\cal G} {\cal P}, \\
{\cal P}&=\frac{{\cal G}_{\uparrow}-{\cal G}_{\downarrow}}{{\cal G}_{\uparrow}+{\cal G}_{\downarrow}},\\
{\cal G}_{\phi}&={\cal G}_{\phi,\uparrow}-{\cal G}_{\phi,\downarrow}.
\label{C4}
\end{align}
Here, ${\cal G}_P$ and ${\cal G}$ refer to conductances given in terms of spin-dependent boundary conductances ${\cal G}_{\uparrow,\downarrow}$. A spin-polarized boundary necessarily leads to spin-dependent scattering phases that are accounted for by a parameter ${\cal G}_{\phi}$ which is the most relevant parameter to modify the superconducting correlations. This modification appears in the pair amplitudes $(\mathfrakF,\mathfraktF)$ of the structure. This can be thought of as the ferromagnet imprinting its magnetic correlations to the proximity coupled superconductor in its immediate vicinity which influences the transport properties of the structure. \\

\subsection{Determination of the Green function $\hG_C$ of the central node}
The Green function $\hG_C$ is calculated within Andreev circuit theory and is used to evaluate the ferromagnetic influence on the transport properties of the system through the superconductor  via a self-energy contribution to the Usadel equation.
From the Kirchhoff rule, Eq.~\eqref{EqKirchhoffRule}, the contact Green Function $\hG_C$ in the central node $C$ is determined by solution of the equation
\begin{align}
[\h{M}(x,\epsilon),\hG_C(x,\epsilon)]=\hat 0
\label{EqContactGreenfunction}
\end{align}
where $\h{M}(x,\epsilon)$ is given by
\begin{align}
\h{M}(x,\epsilon)&=\frac{{\cal G}_q}{4 \epsilon_{\rm Th}}\hG_{\text{Leak}}(\epsilon)+\frac{1}{2}{\cal G} \cdot \hG^{F}+\frac{1}{2}{\cal G}_{P}\cdot \{\h{\kappa}, \hG^{F}\} \nonumber \\
&\qquad -\frac{\pi}{2} {\cal G}_{\phi}\cdot \h{\kappa}+\frac{1}{2}{\cal G}_S \cdot \hG_{0}(x,\epsilon) .
\label{EqMContribution}
\end{align}
%We make the general ansatz
%\begin{align}
%\h{G}_C(x,\epsilon)=\lambda \mathbb{1}_{4x4}+\mu \h{M}(x,\epsilon)  .
%\end{align}
Eq.\! \eqref{EqContactGreenfunction} is supplemented by the normalization condition
\begin{align}
\hG^{2}_C=-\pi^2 \h 1 ,
\label{EqContactGreenfunctionnormalisation}
\end{align}
which means that (a) the matrix $\hG_C$ is diagonalizable and (b)
the only eigenvalues of $\hG_C$ are $\pm i\pi$.
Eq.~\eqref{EqContactGreenfunction} then ensures that if $\h{M}$ is diagonalizable (which in our case holds true), then $\h{M}$ and $\hG_C$ can be diagonalized simultaneously and have a common set of eigenvectors.
Additionally, we demand that the eigenvalues of the contact Green function be continuously connected to those of the normal state. \cite{NazarovYuli,Machon14} 
With these constraints 
%we only have $\lambda =0$ and $\mu=i\pi$ as possible solutions, and 
the Green function $\hG_C$ is written as
\begin{align}
\hG_C(x,\epsilon)=i\pi \h{U}_{M}^{-1}\text{sgn}[\text{Im}(\h{D}_{M})]\h{U}_{M}
\label{EQ:CavityGreenFunction}
\end{align}
with $\h{D}_{M},\h{U}_{M}$ containing the eigenvalues and eigenvectors of matrix $\h{M}$, respectively, and sgn denoting the sign function applied to the imaginary part of each eigenvalue.

\begin{figure*}[t]
\includegraphics[width=8cm]{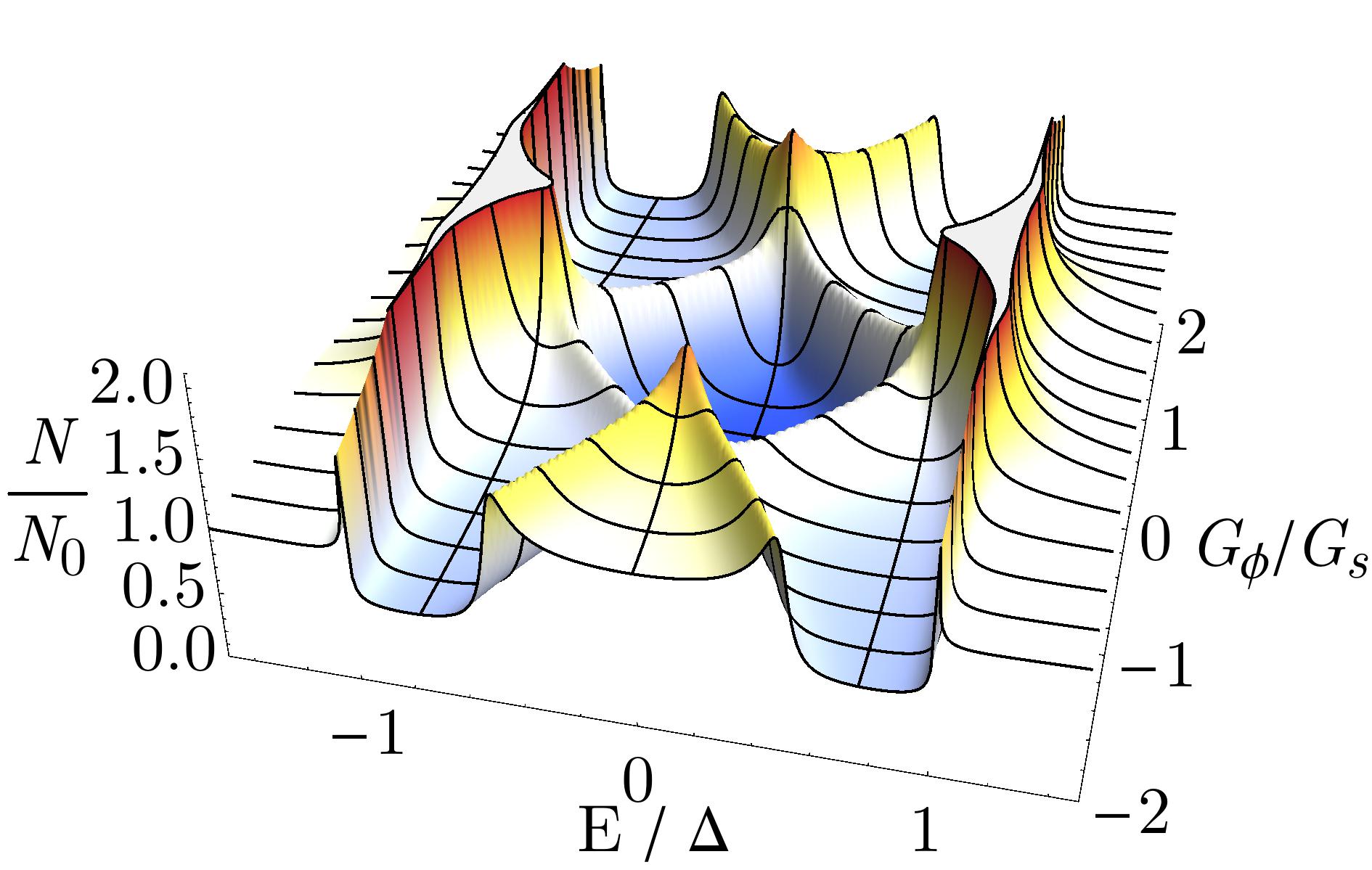}
\includegraphics[width=7cm]{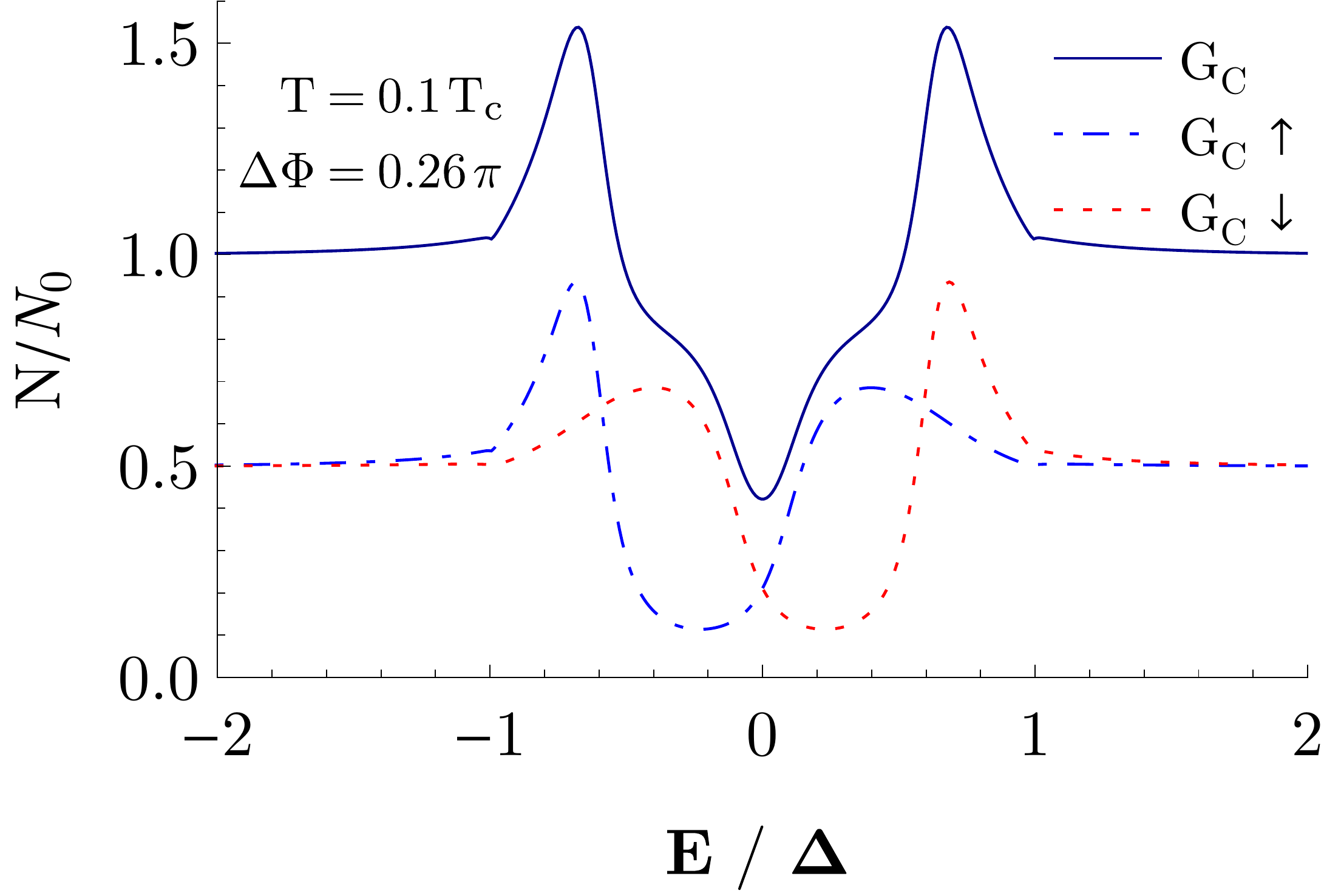}
\caption{\label{FIG:4a} 
Density of states (DOS) $N(E)$ of the contact Green function $\hG_C$ in the central node C (see \Fig{FIG:2}). $N_0$ is the density of states in the normal state at the Fermi energy. 
Left: Dependence on ${\cal G}_{\phi}$ and on $E/\Delta_0 $ calculated for a homogeneous superconductor for $\Delta \Phi=0$.
%The parameter set for this plot is $\{{\cal P}=0.9,{\cal G}=0.1{\cal G}_S, \epsilon_{\rm Th}=0.51\Delta_0\}$, where $\Delta_0=\Delta(T=0)$. 
%A non-zero ${\cal G}_{\phi}$ acts similar to a ferromagnetic exchange field, as it shifts spin-up and spin-down contributions in opposite directions. 
%}
%\end{figure}
%\begin{figure}[t]
%\includegraphics[width=8cm]{{LDOSGC_pre2.3562_T=0.1000_J=0.0_Gphi=0.250_loop=4_phase=0.314_dw=0.00_sx=61_ds=5.0_df=2.0_DeltaE=0.010_GPup=0.045_GPdown=0.000_Gup=0.100_Gdown=0.000_EThouless=0.900_Gs=1.000}.pdf}
%\caption{
\label{FIG:4b} 
Right: 
$N/N_0$ at $x=6\xi$ in the middle of the weak-link S-(S|F)-S structure for $\Delta \Phi=0.26\pi $. In addition, the spin-resolved DOS is shown, to illustrate the spin-splitting due to a non-zero ${\cal G}_{\phi}$.
Parameters are as in Table \ref{Tab:SystemParameters}.}
\end{figure*}

One can now calculate measurable quantities such as the density of states which depends on the set of parameters ${\cal G},{\cal P},{\cal G}_{\phi}$.
The parameter ${\cal G}_{\phi}$ has a similar effect on the density of states as a ferromagnetic exchange field. A non-zero value of ${\cal G}_{\phi}$ spin-splits the density of states in the central node $C$, see 
%figures \ref{FIG:4a} and \ref{FIG:4b}. 
Fig.~\ref{FIG:4a}.
On the left in this figure
we plot as an example the density of states inside the central node $C$ 
by taking the analytic value for a homogeneous superconductor $\hG_0(\epsilon)$ (independent of spatial coordinate $x$). 
%The density of states for $\hG_C$ is evaluated using Eq.\! \eqref{EQ:DOS}. 
This result agrees with the one shown in Ref. \onlinecite{Machon13}.
On the right hand side in Fig.~\ref{FIG:4a} we show a typical example for the density of states inside the central node $C$ at coordinate $x=6\xi $ for the S-(F|S)-S system we consider, for parameters shown in Table \ref{Tab:SystemParameters}.
%It can be seen in \Fig{FIG:4a}.
\begin{table}[b]
\begin{center}
\begin{tabular}{ c | c | c | c | c | c | c}
${\cal G}$ &$\rho_S D/Ad$
%${\cal G}_{F/S}$
& ${\cal G}_{\phi}$& ${\cal P}$ & 
$\epsilon_{\rm Th}/{\cal G}_q$ & $d_s$ & $d_f$\\ \hline
0.1 ${\cal G}_S$ & 
%0.75$\pi$ ${\cal G}_S $
0.75 $\pi$ 
& 0.25 ${\cal G}_S$ &0.9 & 0.51 $\Delta_0 /{\cal G}_S$ & 5.0 $\xi$ & 2.0 $\xi$
\end{tabular}
\caption{
If not explicitly stated otherwise, we calculate all observables of our system for these parameters. 
%The conductance parameters ${\cal G}_S$, ${\cal G}$, ${\cal G}_\phi$, ${\cal P}$ are explained in Fig.~\ref{FIG:2} and in Eqs.~\eqref{C1}-\eqref{C4}.
%Here, ${\cal G}$ denotes the conductance value of a boundary between an $S$ terminal and a node, 
%${\cal G}_{F/S}$ is the conductance value defined in Eq.\! \eqref{EQ:DefinitionofGscC}. 
%$\rho_SD/Ad$ is a parameter that appears in Eq.~\eqref{EQ:DefinitionofSigma},
%${\cal G}_{\phi}$ is the spin-mixing parameter  at the $F/S$ interface, ${\cal P}$ the polarization of the ferromagnet,
%$\epsilon_{\rm Th}$ is the Thouless energy, $\Delta_0=|\Delta(T=0)|$, 
% and $d_s$ and $d_f$ denote the length of the superconducting banks to either side of the ferromagnet and the length of the superconductor/ferromagnet proximity block, respectively, see \Fig{FIG:1}. }
}
\label{Tab:SystemParameters}
\end{center}
\end{table}

\subsection{Implementation of boundary conditions}
In $z$-direction we have two interfaces, an $S/F$ boundary at $z=0$ and an $S/I$ boundary at $z=d$ (see \Fig{setup}), where the Green function is subjected to boundary conditions. We use Nazarov's boundary condition for spin-active \cite{Cottet09,Machon13} and spin-inactive interfaces \cite{NazarovYuli,NazarovYuli2} to define the matrix current $\h{I}_S$ in the superconductor in the vicinity of the $S/F$ and the $S/I$ interface:
\begin{align}
\h{I}_{S}=\mp\frac{A}{\rho_{S}}\hG_{S}(x,z,\epsilon)\left. \frac{\mathrm d}{\mathrm dz} \hG_{S}(x,z,\epsilon)\right|_{z=0,(d)}
\end{align}
where $A$ is the contact area of the boundary and the parameter $\rho_{S}$ denotes the resistivity of the $S$ material and the minus (plus) sign refers to $z=0(d)$.
At the $S/I$ boundary the matrix current $I_{S}$ has to vanish, which in linear order is automatically ensured by the chosen parameterization,
\begin{align}
\h{I}_{S}&=\hG_{0}(x,\epsilon)\left. 2\hG_1(x,\epsilon)(z-d)\right|_{z=d}=\hat 0 \nonumber
.
\end{align}
At the $S/F$ boundary at $z=0$, the matrix current in linear order in $(z-d)$ is
\begin{align}
\h{I}_{S}&=-\frac{A}{\rho_S}\hG_{0}(x,z,\epsilon) 2\hG_1(x,\epsilon)\left.(z-d)\right|_{z=0}\nonumber\\
&=\frac{A \cdot d}{\rho_S}2\hG_{0}(x,\epsilon)\hG_{1}(x,\epsilon).\label{EqNazarovStrom1}
\end{align}
%The general form of a matrix current within Andreev circuit theory can always be cast into the form of a commutator with the Green function $\hG_C$, see Eq.\! \eqref{EQ:GeneralMCurrent} and Eq.\! \eqref{EQ:ScTerminal}; we have made use of this fact in Eq.\! \eqref{EqBoundaryCurrentCommutationRelation}. %
Matrix current conservation\cite{NazarovYuli,NazarovYuli2} requires this expression to be equal to the one in Eq.~\eqref{EQ:ScTerminal}, which leads to
%We identify this current with the current we obtained in Andreev Circuit theory, (Eq.\! \eqref{EQ:ScTerminal}) here repeated for convenience
%\begin{align}
%\h{I}_{S,C}&=\frac{1}{2}{\cal G}_S[\hG_{0}(x,\epsilon),\hG_C(x,\epsilon)] .
%\label{EqBoundaryCurrentCommutationRelation}
%\end{align}
%Matrix current conservation requires $\h{I}_{S,C}=\h{I}_{S}$.
%With this identification we equate Eq.\! \eqref{EqNazarovStrom1} and Eq.\! \eqref{EqBoundaryCurrentCommutationRelation},
\begin{align}
\hG_{0}(x,\epsilon)\hG_{1}(x,\epsilon)&=-\frac{{\cal G}_S \rho_S }{4A\cdot d} [\hG_C(x,\epsilon),\hG_{0}(x,\epsilon)] \label{EQ:G1Contribution}.
\end{align}
Since the contact Green function $\hG_C(x,\epsilon)$ is known from the Kirchhoff rule Eq.\! \eqref{EqContactGreenfunction} and $\hG_{0}(x,\epsilon)$ is known from the solution of the Usadel equation, this equation determines the perturbation that is defined in Eq.~\eqref{EqNewUsadelEquationWithSigma} and Eq.~\eqref{EqUsadelwithContribution}, as
\begin{align}
\h{\Sigma}(x,\epsilon)&\equiv \frac{1}{2 \pi}
\frac{\rho_S D }{A \cdot d} 
%{\cal G}_{F/S} 
{\cal G}_S\cdot \hG_C(x,\epsilon) \label{EQ:DefinitionofSigma} .
%\\
%{\cal G}_{F/S}&\equiv \rho_S \frac{D}{A\cdot d} {\cal G}_S .
%\label{EQ:DefinitionofGscC}
\end{align}
\begin{figure*}[t]
\includegraphics[width=8cm]{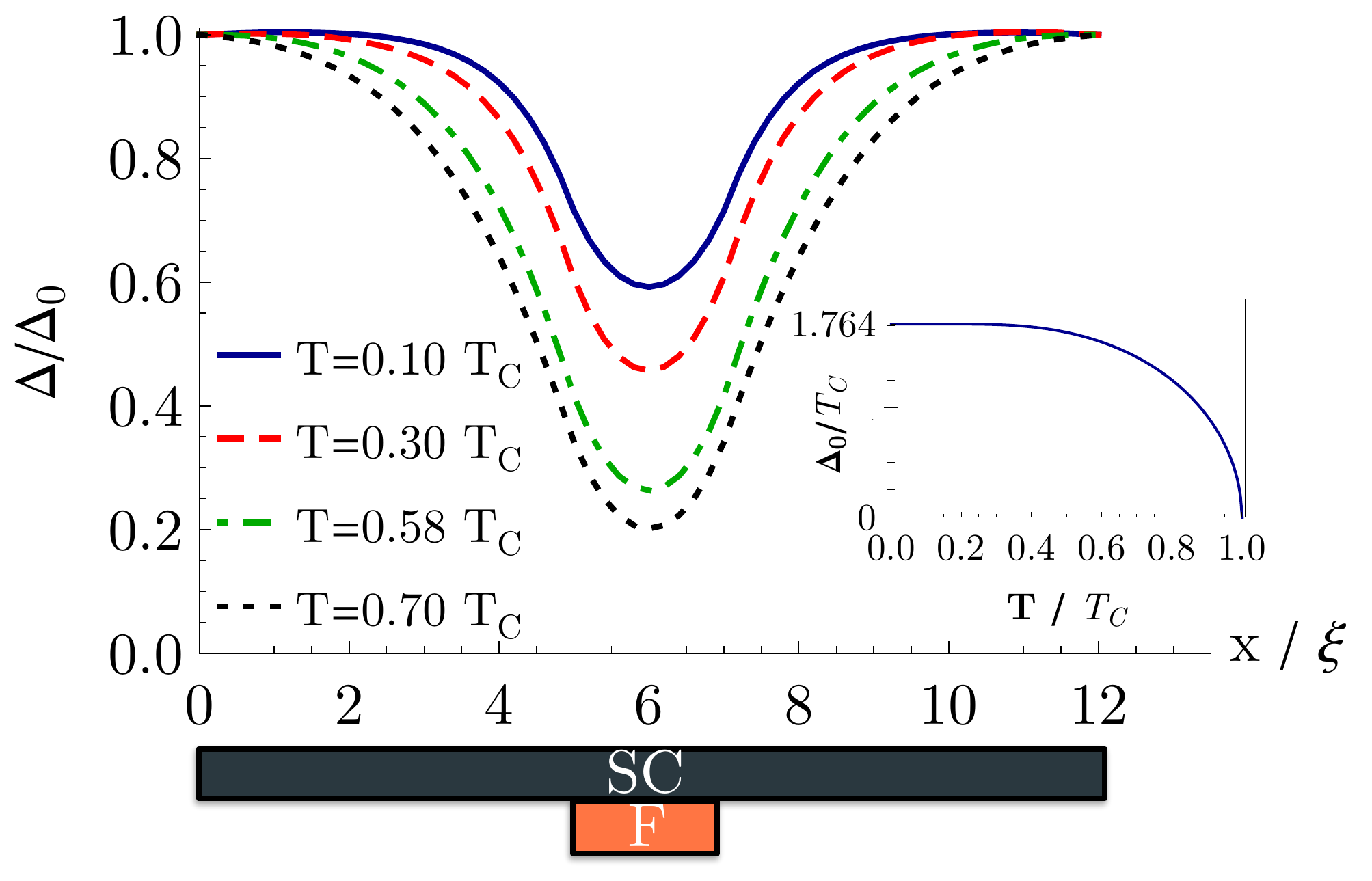}
\includegraphics[width=8cm]{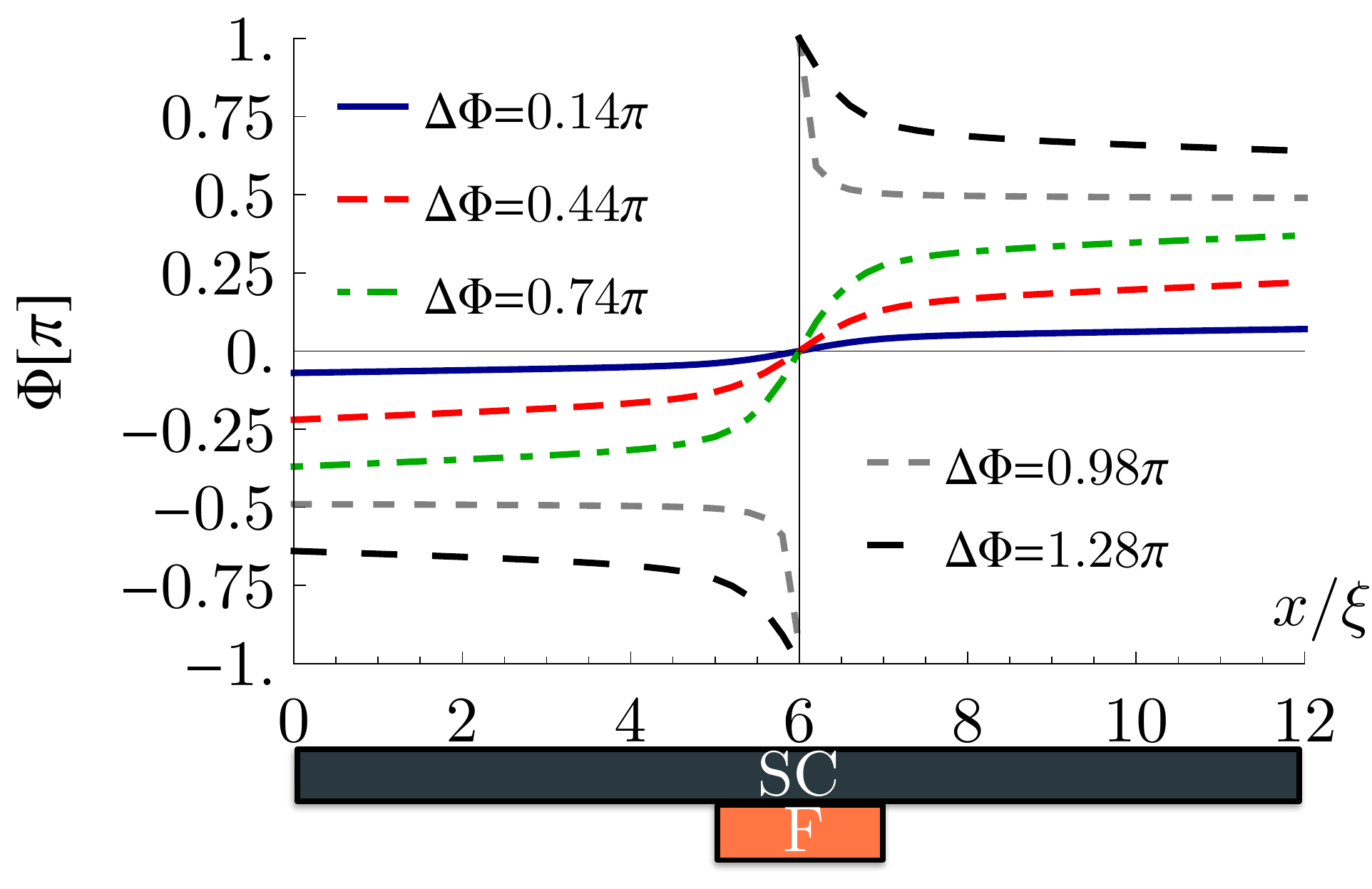}
\caption{\label{FIG:5a} 
Left: Self-consistently calculated pair potential with Eq.\! \eqref{EqSelfConsistencyEquation} for various temperatures with small external phase difference $\Delta \Phi=0.26\pi$. Each Plot has been normalized to the homogeneous value $\Delta_0(T)$ such that all curves are normalized. The pair potential is suppressed in the ferromagnetic region that is located at $x \in (5,7)\xi$. Inset: Temperature-dependence of the pair potential for a homogeneous BCS-superconductor.
%}
%\end{figure}
%\begin{figure}[t]
%\includegraphics[width=8cm]{{SSFSPhase3}.pdf}
%\caption{
\label{FIG:5b} 
Right: Phase evolution across the system for different applied phase differences at a temperature of $T=0.58 T_C$} .
\end{figure*}
We summarize the iterative procedure to self-consistently calculate the pair potential $\Delta$ and the Green function $\hG_C$ below: 
\begin{enumerate}
\item Numerically solve the Usadel equation Eq.\! \eqref{EqNewUsadelEquationWithSigma} to obtain $\hG_0(x,\epsilon) \quad \forall (x,\epsilon)$
\item Application of the Kirchhoff rules $\sum_{i}I_{i}=0$ leads to the Green function $\hG_C(x,\epsilon)$. 
\item Application of spin-conserving and spin-dependent boundary conditions show that the Green function $\hG_C$ determines a self-energy contribution to the Usadel equation written as $\hat{\Sigma}(x,\epsilon)$
\item Solve the Usadel equation with the new self-energy contribution $\hat{\Sigma}(x,\epsilon)$
\item Calculate the order-parameter $\DDelta(x)=\Delta_{0}(x)e^{i\Phi(x)} \cdot i\sigma_y$ by solving the mean-field self-consistency equation Eq.\! \eqref{EqSelfConsistencyEquation}
\item Repeat the iteration procedure until the order-parameter satisfies a convergence criterion.
\end{enumerate}
Note that in the iteration cycle the self-energy $\hat{\Sigma}(x,\epsilon)$ as well as the order-parameter $\DDelta(x,T)$ vary simultaneously. The self-consistent iteration cycle is repeated until both of them have converged.

%%%%%%%%%%%%%%%%%%%%%%%%%%%%%%%%%%%%%%%%%%%%%%%
\section{Observables}
\label{Sec:Observables}
If not explicitly stated otherwise, we calculate all observables of our system for the parameters that are given in Table \ref{Tab:SystemParameters}.
We apply a finite phase difference $\Delta \Phi\equiv \Phi(x=L)-\Phi(x=0)$ to the outer superconducting electrodes, which gives rise to a Josephson current through the weak link.
Formally, this is introduced by the substitution:
\begin{align}
\Delta(x,T) &\to \Delta_0(x,T)e^{i\Phi(x)}, \quad \Phi(x) \in \mathbbm{R} ,\\
\Phi(x=&\text{Left border})=-\frac{\Delta\Phi}{2},\\
\Phi(x=&\text{Right border})=\frac{\Delta\Phi}{2} .
\end{align}
\subsection{Pair potential and phase evolution}
\label{Sec:PairpotentialCalculation}
The pair potential $\DDelta(x,T)$ is calculated by solving a self-consistency equation. 
Here, we are concerned with the following $4 \times 4$ matrix structure arising from particle-hole and spin degrees of freedom:
\begin{align}
\h{\Delta}( x,T)=\left(
\begin{array}{cc}
 0 & \DDelta( x,T) \\
 \DDelta^*( x,T) & 0 \\
\end{array}
\right)
\end{align}
where $\DDelta( x,T)=\Delta( x,T) i\sigma_y $.\\
The self-consistency equation reads 
\begin{align}
\DDelta( x,T)=
\lim_{\epsilon_c\to\infty }
\frac{\frac{1}{2\pi i}\int\limits_{-\epsilon_c}^{\epsilon_c} d\epsilon \tanh\l \epsilon/2T\r \mathfrakF\l x,\epsilon,T\r}{\int\limits_{-\epsilon_c}^{\epsilon_c}\frac{d\epsilon}{2\epsilon}\tanh\l\epsilon /2 T\r+\text{ln}(T/T_c)} .
\label{EqSelfConsistencyEquation}
\end{align}
The equation is numerically evaluated with a sufficiently large, temperature-independent, energy cut-off $\epsilon_{c}$.
Apart from the usual suppression of superconductivity with increasing temperature, the pair potential is strongly suppressed in the region of the ferromagnet, see on the left hand side of \Fig{FIG:5a}. 
When applying a fixed phase difference $\Delta \Phi$ to the outer superconducting electrodes, the spatial evolution of the phase is determined by the self-consistency equation Eq.\! \eqref{EqSelfConsistencyEquation}.
The self-consistent evolution of the phase across the system is shown on the right hand side of \Fig{FIG:5b}. In the numerical iteration process, we fix the phase difference as well as the absolute value of the pair potential at the left and the right-hand side of our structure. The latter is
given by the well-known temperature-dependence of a homogeneous BCS-type superconductor (see inset, \Fig{FIG:5a}) 
%%%%%%%%%%%%%%%%%%%%%%%%%%%%%%%%%%%%%
\label{Sec:Density of states}
\subsection{(Local) Density of states}
\begin{figure}[b]
\includegraphics[width=8cm]{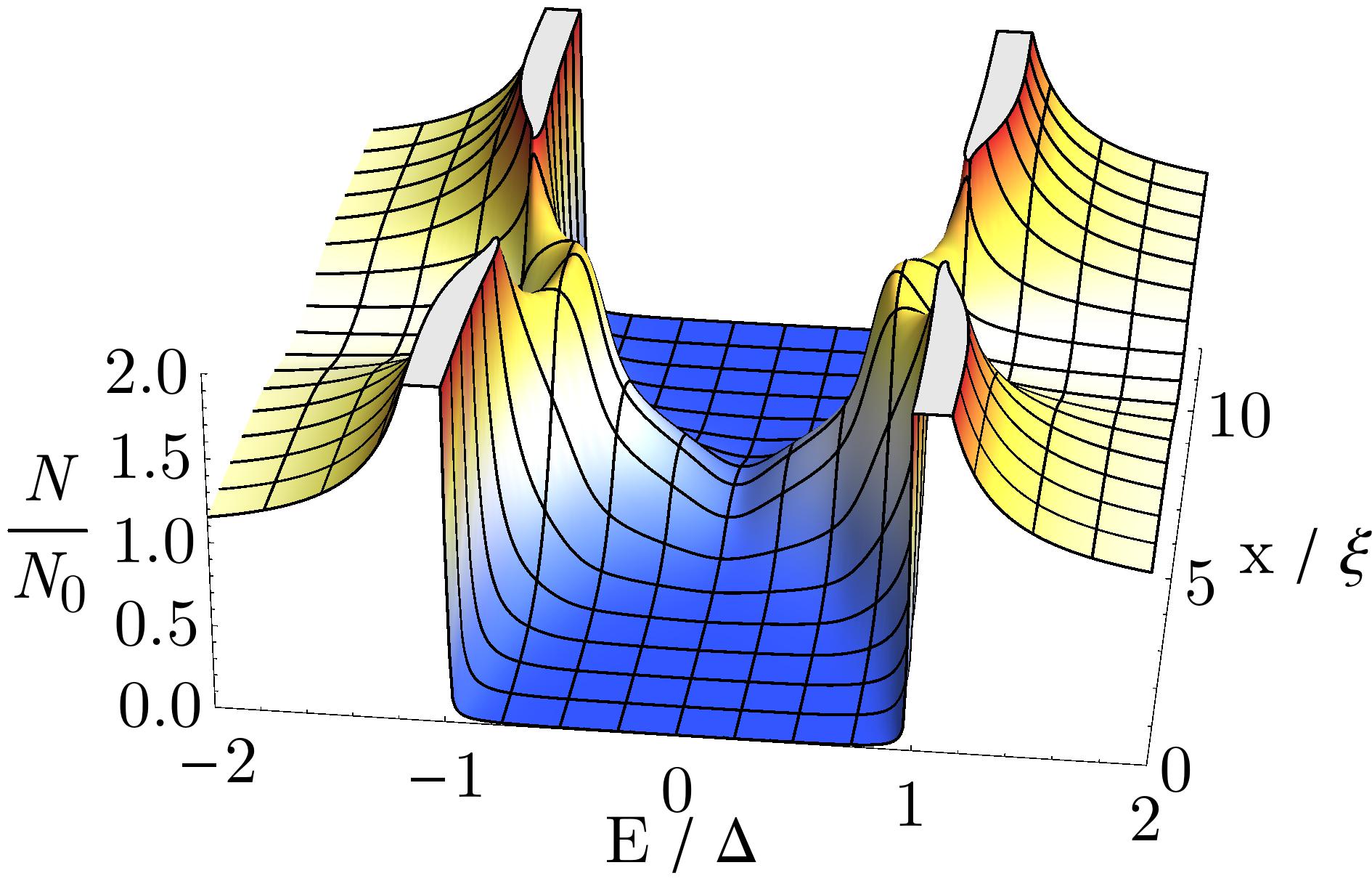}
\caption{\label{FIG:6} Density of states as a function of energy $E$ and spatial coordinate $x$. The (S|F) weak link extends from $x=5\xi $ to $7\xi$. 
The DOS is calculated for $\Delta \Phi=0.26 \pi$, at $T=0.1 T_C$.}
\end{figure}

The local density of states for the system is given by
\begin{align}
N(x,\epsilon)=-\frac{N_0}{2\pi}\text{Im}\left(\text{Tr}_2[\mathfrakG(x,\epsilon)]\right) .
\label{EQ:DOS}
\end{align}
In Fig.~\ref{FIG:6} the spatial variation of the local density of states is shown.
\begin{figure*}[t]
\includegraphics[width=7.5cm]{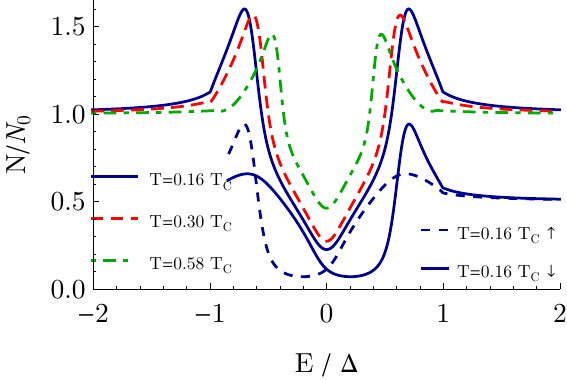}
\includegraphics[width=8cm]{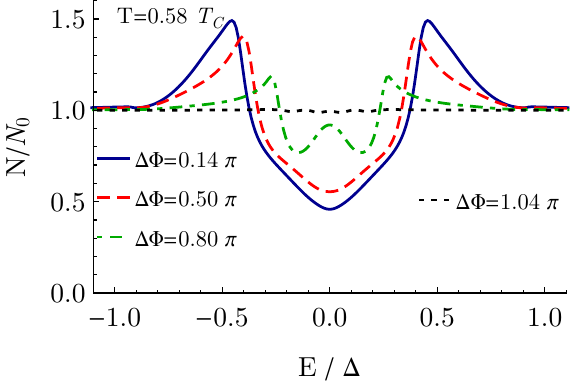}
\caption{Left: Local density of states at $x=6 \xi$ in the weak-link structure. The LDOS is spin-split into a spin-up and a spin-down contribution due to a non-zero ${\cal G}_{\phi}$ component. The lines corresponding to temperatures $T_{C,\uparrow(\downarrow)}$ refer to the local spin-density of states $N_{\uparrow}/N_0,N_{\downarrow}/N_0$, respectively, with the total density of states defined as $N=N_{\uparrow}+N_{\downarrow}$ and $N_0$ as the density of states at the Fermi energy in the normal state.\label{FIG:7a} 
%}
%\end{figure}
%\begin{figure}[t]
%\includegraphics[width=8cm]{{LDOSStroms2}.pdf}
%\caption{
Right: Variation of the density of states at $x=6\xi$ in the middle of the structure with an applied phase-difference $\Delta\Phi$. 
%An increasing phase difference fills up the density of states with additional Andreev bound states. 
\label{FIG:7b}}
\end{figure*}
It can be seen that the DOS retains its characteristic structure far from the (S|F) weak link. In the (S|F) weak-link region additional subgap Andreev bound states appear. 
These are present also in the superconducting electrodes within a coherence length from the weak-link region, in particular below the gap edges of the bulk density of states.

The local density of states at $x=6\xi$ (the middle of the structure) is shown in 
%Figs.~\ref{FIG:7a} and \ref{FIG:7b}, 
Fig.~\ref{FIG:7a}.
In the middle of the investigated structure, the local density of states for the system shows a proximity induced narrowing in its spectrum that is reminiscent of proximity induced minigaps in S-N-S systems. This narrowing is induced by a spin-split DOS as can be seen in \Fig{FIG:7a}. The presence of the ferromagnet, which is encoded in a non-vanishing spin-mixing parameter ${\cal G}_{\phi}$ shifts spin-up and spin-down contributions to the density of states energetically apart. An additionally applied phase difference to the outer superconducting electrodes leads to a gradual reduction of the gap size. This is due to additional subgap Andreev bound states \cite{Andreev}, which are shifted in the presence of a superflow, see \Fig{FIG:7b} right. A zero-bias peak appears for sufficiently large phase differences. 

\subsection{Spin-Magnetization}
\label{ChapterSpinMagnetisationwithPlots}
We make use of the usual notation to express the normal and anomalous pair amplitudes $\mathfrakG$ and $\mathfrakF$, respectively,
\begin{align}
\mathfrakG&=G_0\mathit{1}+\boldsymbol{G} \cdot \boldsymbol{\sigma}\\
\mathfrakF&=(F_0\mathit{1}+\boldsymbol{F} \cdot \boldsymbol{\sigma})i\sigma_y
.
\end{align}
A general feature of SF-proximity influenced systems is that the presence of non-zero triplet amplitudes $\b{F}$ entails the presence of a non-zero $\b{G}$.\cite{Champel05}
This means that through a non-zero $G_z$, the spin-up and spin-down contribution to the local density of state are not degenerate any more. 
Thus, due to proximity to the ferromagnet, the superconductor develops a spin-magnetization in the vicinity of the S|F interface.\cite{Alexander85,Tokuyasu88,Bergeret04} 
The induced spin-magnetization $\b{m}(x)$ can be calculated in the following way \cite{Eschrig15}
\begin{align}
\b{m}(x)
=&2 N_0 T\int\limits_{-\infty}^{\infty}\frac{d\epsilon}{4\pi i} \quad \left[\b{G}(x,\epsilon+i\delta)\tanh\left(\frac{\epsilon+i\delta}{2T}\right)\right. \nonumber \\
&\left.\qquad-\left(\b{G}(x,\epsilon+i\delta) \right)^{\dagger}\tanh\left(\frac{\epsilon-i\delta}{2T}\right)\right]
\label{EqSpinMagnetisierung}
\end{align}
where 
$N_0$ is the density of states at the Fermi energy. 

\begin{figure}[b]
\includegraphics[width=8cm]{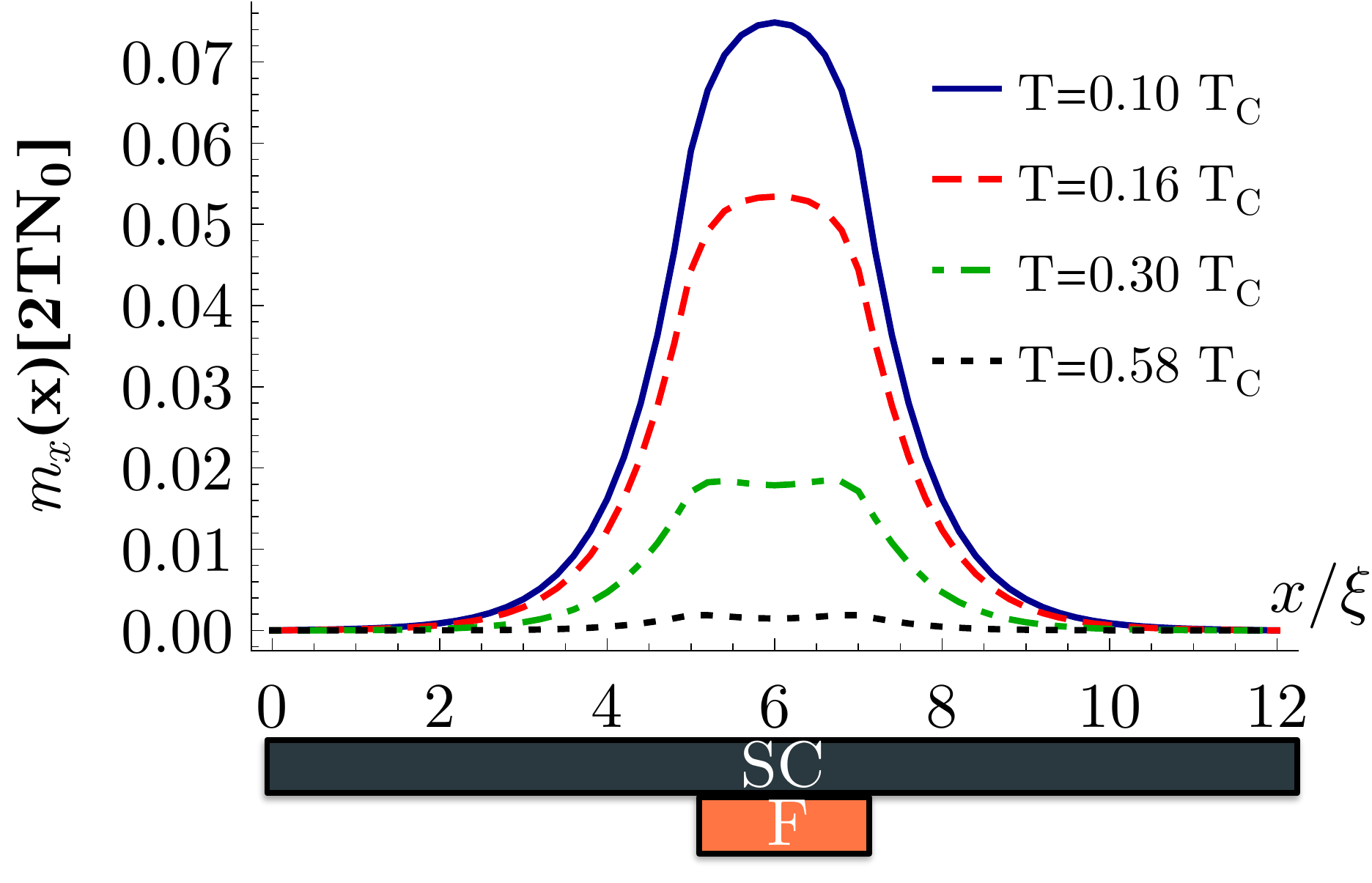}
\caption{\label{FIG:8} Induced spin-magnetization by a non-zero $\b{G}$ component. It can be seen that the system obtains a finite spin-magnetization that is restricted to the contact region of the ferromagnet and the superconductor. Due to the inverse proximity effect and the penetration of the triplet amplitudes in the superconducting region, there is a finite magnetization for $x > 7\xi$ and for $x<5\xi$. The applied phase-difference is $\Delta\Phi=0.26 \pi$.}
\end{figure}
A plot of $m_x(x)$ can be seen in \Fig{FIG:8}.
A non-zero spin-magnetization is induced in the superconducting material that sits on top of the ferromagnetic material, i.e. at\! $x\in (5,7)\xi$. Due to the inverse proximity effect, a non-zero magnetization can penetrate into the adjacent superconducting blocks as it is indicated in \Fig{FIG:8}.
The calculations show that the ferromagnet imprints its magnetic structure onto the superconductor.
\subsection{Weak-link Current-Phase Relationships}
\label{Sec:JosephsonCurrents}

A finite phase difference $\Delta \Phi \neq 0$ gives rise to a Josephson current through the system.
The Josephson currents themselves are spatially conserved, $\partial_{x}J(x)=0$, only if the system fulfills the self-consistency equations for the pair potential, see Eq.\! \eqref{EqSelfConsistencyEquation}. The current conservation in the presence of the self-energy correction to the Usadel Equation can be shown analytically (see section \ref{Sec:CurrCons}). 
The total current through the superconducting leads is denoted as $J(x)$ and is evaluated as
\begin{align}
&J= 
%-eN_0\int\limits_{-\infty}^{+\infty} \frac{d\epsilon}{4 \pi} \frac{D}{\pi} \frac{\text{Tr}_4}{2}\left(\h{\tau}_3  \left[\hG_0(x,\epsilon) \partial_x \hG_0(x,\epsilon) \right]^{K}\right) \\ \nonumber & =
-eN_0\int\limits_{-\infty}^{+\infty} \frac{d\epsilon}{2 \pi} \frac{D}{\pi} \frac{\text{Tr}_4}{2}\left(\h{\tau}_3 {\rm Re}\left[\hG_0 \partial_x \hG_0 \right]\right) 
\tanh\frac{\epsilon}{2T}
%F(\epsilon) 
\\ \nonumber
&=\frac{\sigma_{N}}{2e}\int\limits_{-\infty}^{+\infty}d\epsilon\text{Tr}_2 \left[
 \text{Re} \l \left\{(\mathit{1}-\gamma\tilde{\gamma})^{-1}\partial_x\gamma(\mathit{1}-\tilde{\gamma}\gamma)^{-1},\tilde{\gamma}\right\}_{+} \right. \right. \nonumber\\& \left. \left.\qquad -
 \left\{(\mathit{1}-\tilde{\gamma}\gamma)^{-1}\partial_x\tilde{\gamma}(\mathit{1}-\gamma\tilde{\gamma})^{-1},\gamma\right\}_{+}\r \right]
\tanh\frac{\epsilon}{2T}
%F(\epsilon)
\end{align}
where 
%$F(\epsilon)=\tanh\l\epsilon/2T\r$ and 
$\sigma_N=e^2N_0D$ is the conductivity in a normal metal with diffusion coefficient $D$ and the density of states at the Fermi level $N_0$. Here $\{A,B\}_{+}=AB+BA$, and
in the last line we have written the current in terms of the Riccati amplitudes.\\
The critical current is the maximum current that is realized in the system as function of phase differences $\Delta \Phi$,
\begin{align}
J_s(T)=\text{\rm max}_{\Delta \Phi }\left[J(\Delta \Phi,T)\right].
\label{Eq:Critical_Current}
\end{align}
%%%%%%%%%%%%%%%%%%%%%%%%%%%%%%%%%%%%%%%
\begin{figure}[t]
\includegraphics[width=8cm]{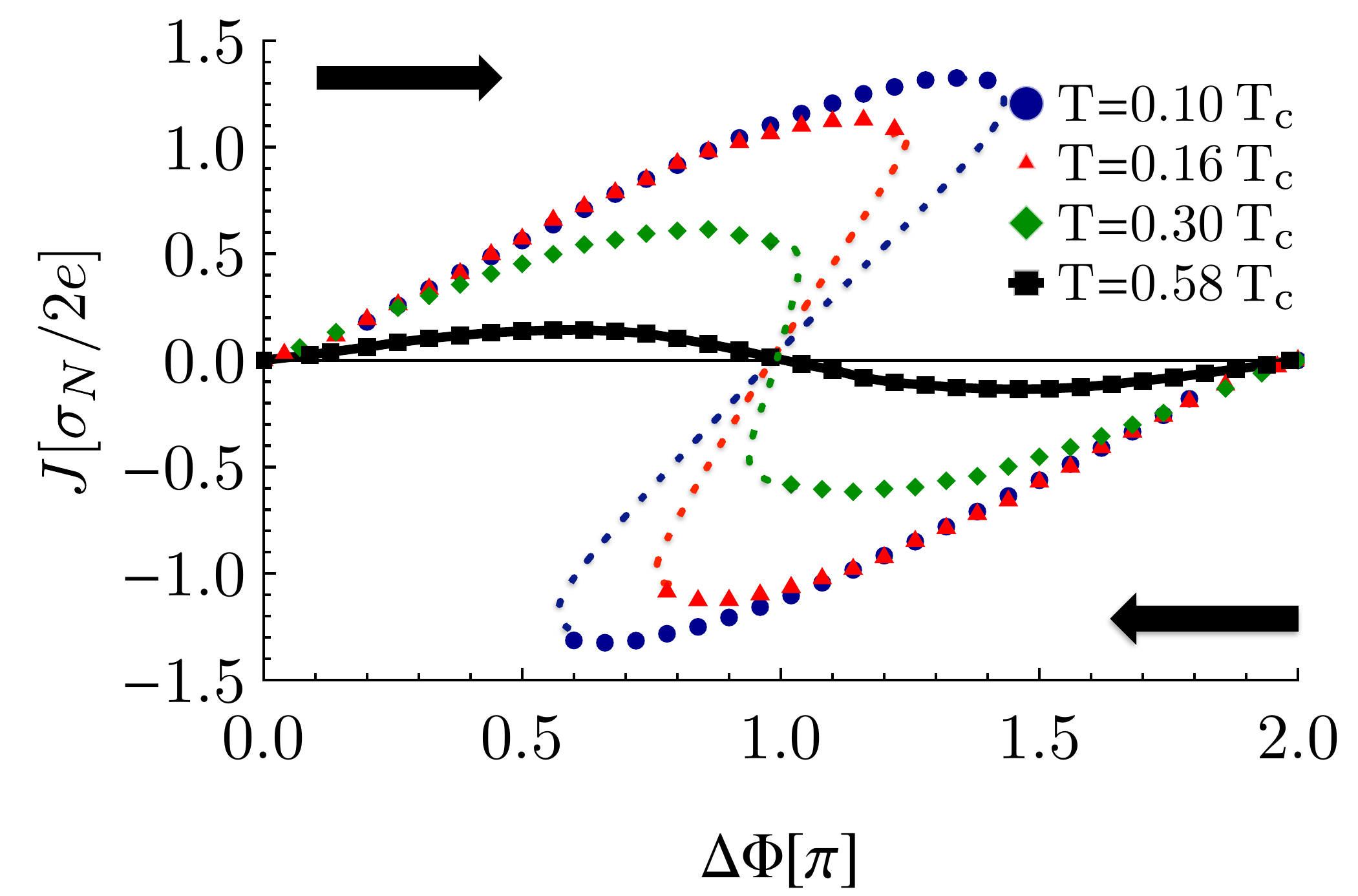}
\caption{\label{FIG:9} Crossover from a multi-valued to a sinusoidal current as the temperature is increased. As indicated by the arrows, the graphs from the upper half are calculated for an increasing phase difference, the graphs from the lower half for a decreasing phase difference. The dotted lines are schematic and indicate the most likely continuation of the current-phase relationship, as the current-phase relationship becomes unstable shortly after the maximum is reached as multiple solutions for the current become possible. We show here only the stable, converged solutions.}
\end{figure}
%%%%%%%%%%%%%%%%%%%%%%%%%%%%%%%%%%%%%%%%%%%%%%
\begin{figure*}
\includegraphics[width=7.5cm]{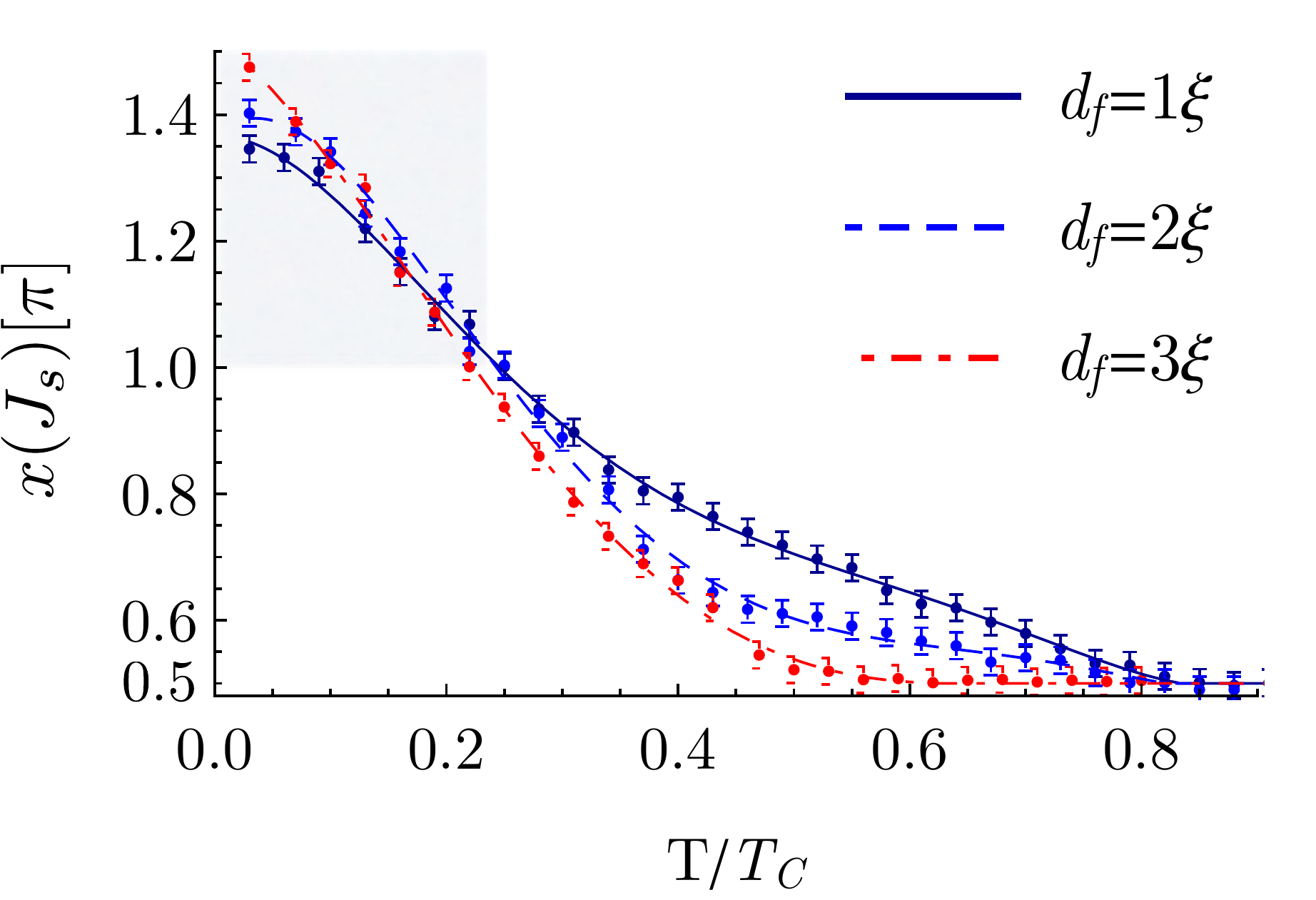}
\includegraphics[width=7.5cm]{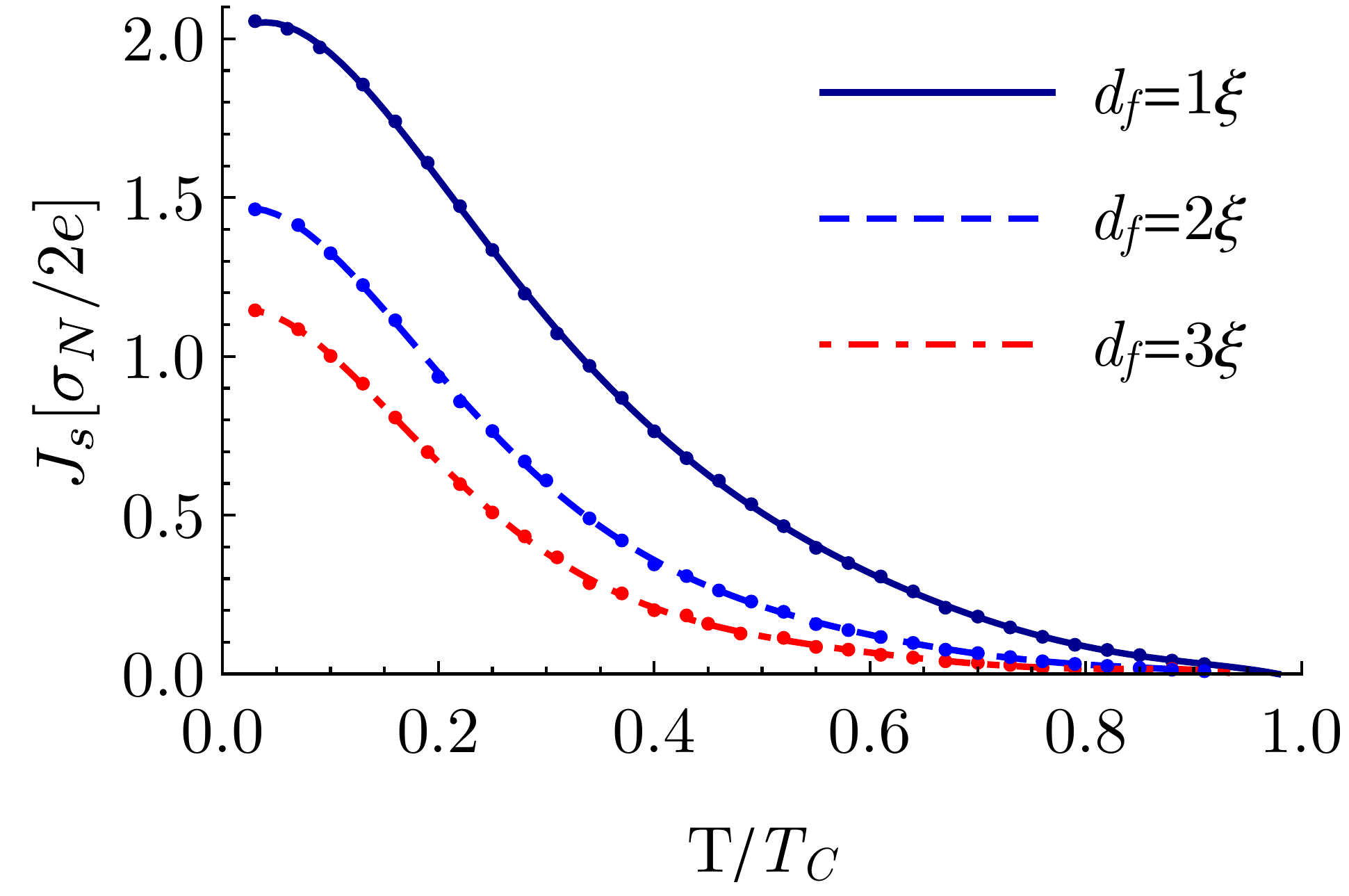}
\caption{\label{Fig:PositionofMaxCurrents} 
Left: Dots represent the superconducting phase difference 
at which the maximal (critical) current occurs as function of temperature. At high temperatures the critical current appears at a value of $\pi/2$ (approximately sinusoidal current-phase relationship), whereas for low temperatures the current-phase relationship is non-sinusoidal. The shaded area indicates the range where the critical current is achieved for $\Delta \Phi > \pi$. 
The error bars indicate the numerical uncertainty in determining the extrema in the current-phase relationship. 
%}
\label{FIG:Critical_Current_Phase_Relationship}
%\end{figure}
%\begin{figure}[b]
%\includegraphics[width=7.5cm]{{TotalJosephsonCurrents25January}.pdf}
%\caption{
\label{Fig:CriticalJosephsonCurrents} 
Right: Critical current for different system sizes. The current is determined by the maximum of the current-phase relationship Eq.\! \eqref{Eq:Critical_Current}. In both plots lines are guides to the eye.
}
\end{figure*}
%%%%%%%%%%%%%%%%%%%%%%%%%%%%%%%%%%%%%%%%%%%%%%%%
The phase-dependence for the weak link structure (S-(F$|$S)-S) in general strongly deviates from a sinusoidal relation at low temperatures. 
In particular, it can become multi-valued in certain ranges of the phase difference $\Delta \Phi $.\cite{Golubov04} If the current-phase relation is single-valued, then
for symmetry reasons $J(\Delta \Phi = \pi)=0$. In the case of a multi-valued current-phase relation it is however possible that the current reaches its maximum value for phases $\Delta \Phi > \pi$, before the current jumps to its negative branch at a critical value of $\Delta \Phi$.

Typical current-phase relationships are depicted in \Fig{FIG:9}. At low temperatures, shortly after the current reaches its maximum value, the system can occupy two different Josephson states that only slightly differ in energy. For the numerical calculation it means that the system can oscillate between the two solutions which makes it numerically difficult to converge to a solution. Stable, converged solutions are therefore shown as symbols, whereas
dotted lines show the most likely continuation for the unstable branches.
For increasing temperature, the current-phase relation approaches a sinusoidal form. The maximum of the current is shifted to lower values of $\Delta \Phi$ as the temperature is increased until they reach a value of $\Delta \Phi=0.5\pi$. 
We track numerically the temperature-dependence of the phase $\chi(J_s)$ where the critical current is reached, and depict the result in \Fig{Fig:CriticalJosephsonCurrents} (left), where it can be seen that for all three lengths of the ferromagnet, critical currents are reached for phase differences $\Delta\Phi>\pi$ at low temperatures.
A temperature-dependence of the critical currents for different lengths of the ferromagnetic block is shown in \Fig{Fig:CriticalJosephsonCurrents} (right).
As expected, we observe that shorter weak links generally increase the critical currents. 

%%%%%%%%%%%%%%%%%
\section{Current Conservation}
\label{Sec:CurrCons}
The Usadel equation entails a spatial conservation law for the Josephson currents. Here, we review that the current is spatially conserved in the presence of the effective self-energy contribution $\hat{\Sigma}(x,\epsilon)$. 
For this we introduce Keldysh matrices
\begin{equation}
\cG_0= \left( \begin{array}{cc}
\hG^R_0 & \hG^K_0 \\ \h 0 & \hG^A_0 \end{array}\right),
\quad 
\hG^K_0=
\left(
\begin{array}{cc}
\;\; \mathfrakG^K & \;\; \mathfrakF^K \\
 -\mathfraktF^K & -\mathfraktG^K \\
\end{array}
\right)
\label{KM}
\end{equation}
where $R$, $A$, and $K$ refer to retarded, advanced, and Keldysh components, respectively.
Similarly, we have
\begin{equation}
\check \Delta= \left( \begin{array}{cc}
\h \Delta & \h 0 \\ \h 0 & \h \Delta \end{array}\right) , \quad
\check \Sigma=\left( \begin{array}{cc}
\h \Sigma^R& \h \Sigma^K \\ \h 0 & \h \Sigma^A \end{array}\right).
\end{equation}
%where $\check 0$ and $\check 1$ are 8$\times$8 zero and unit matrices, respectively.
We also define $\check \tau_3=\mathbbm{1}_{2\times 2} \otimes \h \tau_3 $.
The Usadel equation reads:
\begin{align}
[\epsilon \check{\tau}_3-\check{\Delta}-\check{\Sigma},\cG_{0}]+\frac{D}{\pi}\partial_x[\cG_{0}\partial_x\cG_{0}]=\check 0 
\label{UE}
\end{align}
where $\check 0$ is an 8$\times $8 zero matrix.
Furthermore, the normalization condition generalizes to
\begin{align}
\cG_0^2=-\pi^2 \check{1}
\label{Knorm}
\end{align}
where $\check 1\equiv \mathbbm{1}_{8\times 8}$ is the 8$\times $8 unit matrix.
This condition is very powerful, as it means that the Keldysh matrix $\cG_0$ in Eq.~\eqref{KM} is diagonalizable and its only eigenvalues are $\pm i \pi $.

To derive a current conservation law from the Usadel equation, one has to 
express the physical current in terms of the Green functions,
\begin{align}
J(x)= -eN_0\int\limits_{-\infty}^{+\infty} \frac{d\epsilon}{4 \pi} \frac{D}{\pi} \frac{\text{Tr}_4}{2}\left(\h{\tau}_3  \left[\cG_{0}(x,\epsilon) \partial_x \cG_{0}(x,\epsilon)\right]^{K}\right) 
\end{align}
where Tr$_4$ is a trace over particle-hole and spin space, $e=-|e|$ the charge of the electron, and $N_0$ the density of states per spin at the Fermi level in the normal state.
The Usadel equation \eqref{UE} then leads to
\begin{align}
-eN_0\int\limits_{-\infty}^{+\infty} \frac{d\epsilon}{4 \pi} \frac{\text{Tr}_4}{2}\left(\hat{\tau}_3 [\epsilon \check{\tau}_3-\check{\Delta}-\check{\Sigma},\cG_{0}]^{K}\right)+\partial_x J=0 .
\label{currc}
\end{align}
Under cyclic invariance of the trace, the term involving $\epsilon \check \tau_3 $ vanishes immediately:
\begin{align}
\epsilon \frac{\text{Tr}_4}{2}\left(\left[\cG_{0}(x,\epsilon) -\check\tau_3\cG_{0}(x,\epsilon)\check\tau_3\right]^{K}\right)=0 \label{EQ:SelfConsistency-Equation} .
\end{align}
The second term, involving $\check \Delta $, vanishes only when $\DDelta $ fulfills the self-consistency equation
\begin{equation}
\DDelta( x,T)=\lim_{\epsilon_c\to\infty }\frac{\frac{1}{4\pi i}\int\limits_{-\epsilon_c}^{\epsilon_c} d\epsilon \; \mathfrakF^K\l x,\epsilon,T\r}{\int\limits_{-\epsilon_c}^{\epsilon_c}\frac{d\epsilon}{2\epsilon}\tanh\l\epsilon /2 T\r+\text{ln}(T/T_c)}.
\label{EqSelfConsistencyEqaution2}
\end{equation}
The corresponding contribution in Eq. \eqref{currc}
then reads
\begin{align}
&\int\limits_{-\infty}^{+\infty} \frac{d\epsilon}{4 \pi} \frac{\text{Tr}_4}{2}\left(\hat{\tau}_3 [\check{\Delta},\cG_{0}(x,\epsilon)]^{K}\right) \nonumber \\
&=\lim_{\epsilon_c\to\infty }\int\limits_{-\epsilon_c}^{+\epsilon_c} \frac{d\epsilon}{4 \pi} \frac{\text{Tr}_2}{2}\left(\{\mathfrakF^{K},\DDelta^{*}\}+\{\DDelta,\mathfraktF^{K}\}\right)=0
\end{align}
where Tr$_2$ is a trace over spin, and
Eq.\! \eqref{EqSelfConsistencyEqaution2} was substituted for $\DDelta $ and $\DDelta^\ast$, as well as
\begin{align}
\mathfraktF^{K}( x,\epsilon,T)=
\left(\mathfrakF^{K}( x,-\epsilon,T)\right)^{*} 
= -\left(\mathfrakF^{K}( x,\epsilon,T)\right)^\dagger
\end{align}
used.
Due to the cyclic invariance of the trace, the third contribution from the commutator in Eq. \eqref{currc} vanishes as well:
\begin{align}
&\int\limits_{-\infty}^{+\infty} \frac{d\epsilon}{4 \pi} \frac{\text{Tr}_4}{2}\left(\hat{\tau}_3 [\check{\Sigma}(x,\epsilon),\cG_{0}(x,\epsilon)]^{K}\right)=0 .
\label{A9}
\end{align}
To proof this, we note that
the self energy $\check{\Sigma}$ is proportional to $\cG_C$, see Eq.~\eqref{EQ:DefinitionofSigma}, and that, in generalization of Eq.~\eqref{EqContactGreenfunction}, 
%$\cG_C$ commutes with $\check M$, 
\begin{align}
\left[ \check M (x,\epsilon ), \cG_C (x,\epsilon )\right]=\check 0,
\end{align}
where (see Eq.~\eqref{EqMContribution})
\begin{align}
&\check{M}(x,\epsilon )=\left(\frac{{\cal G}_q}{4 \epsilon_{\rm Th}}\cG_{\text{Leak}}(\epsilon)+\frac{1}{2}{\cal G} \cG^{F} \nonumber\right. \\
&\left. \qquad +\frac{1}{2}{\cal G}_{P}\{\check{\kappa}, \cG^{F}\} -\frac{\pi}{2} {\cal G}_{\phi}\check{\kappa}+\frac{1}{2}{\cal G}_S \cG_{0}(x,\epsilon)\right)
%&\check{\Sigma}=\frac{D}{2 \pi}\frac{{\cal G}_S}{A\cdot d}\rho_S\left(\frac{{\cal G}_q}{4 \epsilon_{\rm Th}}\cG_{\text{Leak}}(\epsilon)+\frac{1}{2}{\cal G} \cG^{F} \nonumber\right. \\
%&\left. \qquad +\frac{1}{2}{\cal G}_{P}\{\check{\kappa}, \cG^{F}\} -\frac{\pi}{2} {\cal G}_{\phi}\check{\kappa}+\frac{1}{2}{\cal G}_S \cG_{0}(x,\epsilon)\right)
\label{A12}
\end{align}
where $\cG_{\text{Leak}}=\mathbbm{1}_{2\times 2} \otimes \hG_{\text{Leak}}$, $\cG^F = \mathbbm{1}_{2\times 2} \otimes \hG^F $, and $\check{\kappa}= \mathbbm{1}_{2\times 2} \otimes \h{\kappa}$,
which all three commute with $\check \tau_3$.
%This means, that $\check {G}_C$ and $\check M$ can be
We will assume that $\check M$ has distinct eigenvalues (if not, we can always add an infinitesimal term to make them distinct; in fact, it suffices that each characteristic value occurs in only one Jordan block in the Jordan normal form of the matrix). 
%In this case, $\check M $ is diagonalizable with a basis of eigenvectors, and furthermore, as $\cG_C$ commutes with $\check M$, both share this basis of eigenvectors. 
Then, a well-known mathematical theorem ascertains that $\cG_C $ can be written uniquely as a polynomial in $\check M$ of at most degree 7 (for 8$\times $8 matrices).
Consequently, we can expand $\check {G}_C$ in the following way,
\begin{align}
\cG_C=\lambda \check 1+\mu \check{M} + \nu \check{M}^2 + \rho \check{M}^3 + \cdots
\label{expand}
\end{align}
%Here, we have, see Eq.\! \eqref{EqMContribution}:
%The coefficients $\lambda $, $\mu $, $\nu $, and $\rho $ etc follow from the linear system of equations
%(see Eq.~\eqref{EQ:CavityGreenFunction})
%\begin{align}
%\mbox{sgn}\big[\mbox{Im} \check D_M\big]=\lambda \check 1+\mu \check{D}_M + \nu \check{D}_M^2 + \rho \check{D}_M^3 + \cdots .
%\end{align}
As $\check M$ is of the form $\check M=\check D+\alpha \cG_0$ where $\check D$ commutes with $\check \tau_3$, and because of the condition \eqref{Knorm}, any power of $\check M$ will only have terms that are of the form $\check D^n$ or $\check D^n\cG_0\check D^m+\check D^m \cG_0 \check D^n$ or $\sum_{{\cal P}[nmk]}\check D^n\cG_0\check D^m\cG_0\check D^k$ etc. (up to maximally terms containing four $\cG_0$'s), where ${\cal P}[nmk]$ means a permutation of $[nmk]$, and $n$, $m$, and $k$ are integers $\ge 0$.
The cyclic property of the trace together with condition \eqref{Knorm} then leads to vanishing contributions for each term in Eq.~\eqref{expand} when introduced into Eq.~\eqref{A9} using Eq.~\eqref{A12}. 
For example,
\begin{align}
\check\tau_3\sum_{{\cal P}[nmk]}(\check D^n\cG_0\check D^m\cG_0\check D^k \cdot \cG_0-\cG_0 \cdot \check D^n\cG_0\check D^m\cG_0\check D^k)
\nonumber
\end{align}
turns, when using cyclic permutation under the trace and commutation between $\check D$ and $\check \tau_3 $, into
\begin{align}
\check \tau_3\sum_{{\cal P}[nmk]}(\check D^n\cG_0\check D^m\cG_0\check D^k \cG_0-\check D^k \cG_0 \check D^n\cG_0\check D^m\cG_0), 
\nonumber
\end{align}
and as $[knm]$ is a permutation of $[nmk]$, the two terms cancel when summing over all permutations
(we use $[\check \tau_3 \check A]^K=\h \tau_3[\check A]^K\equiv \h \tau_3 \h A^K$), 

Consequently, collecting all results together, it follows that for our theory
\begin{align}
\partial_x J(x)=0,
\end{align}
which is the (stationary) charge conservation law.
%%%%%%%%%%%%%%%%%%%%%%%%%%%%%%%%%
\section{S-F-S structure with a magnetic domain wall}
\label{Sec:SFSMagneticDomainWall}
\begin{figure}[t]
\includegraphics[width=5.5cm]{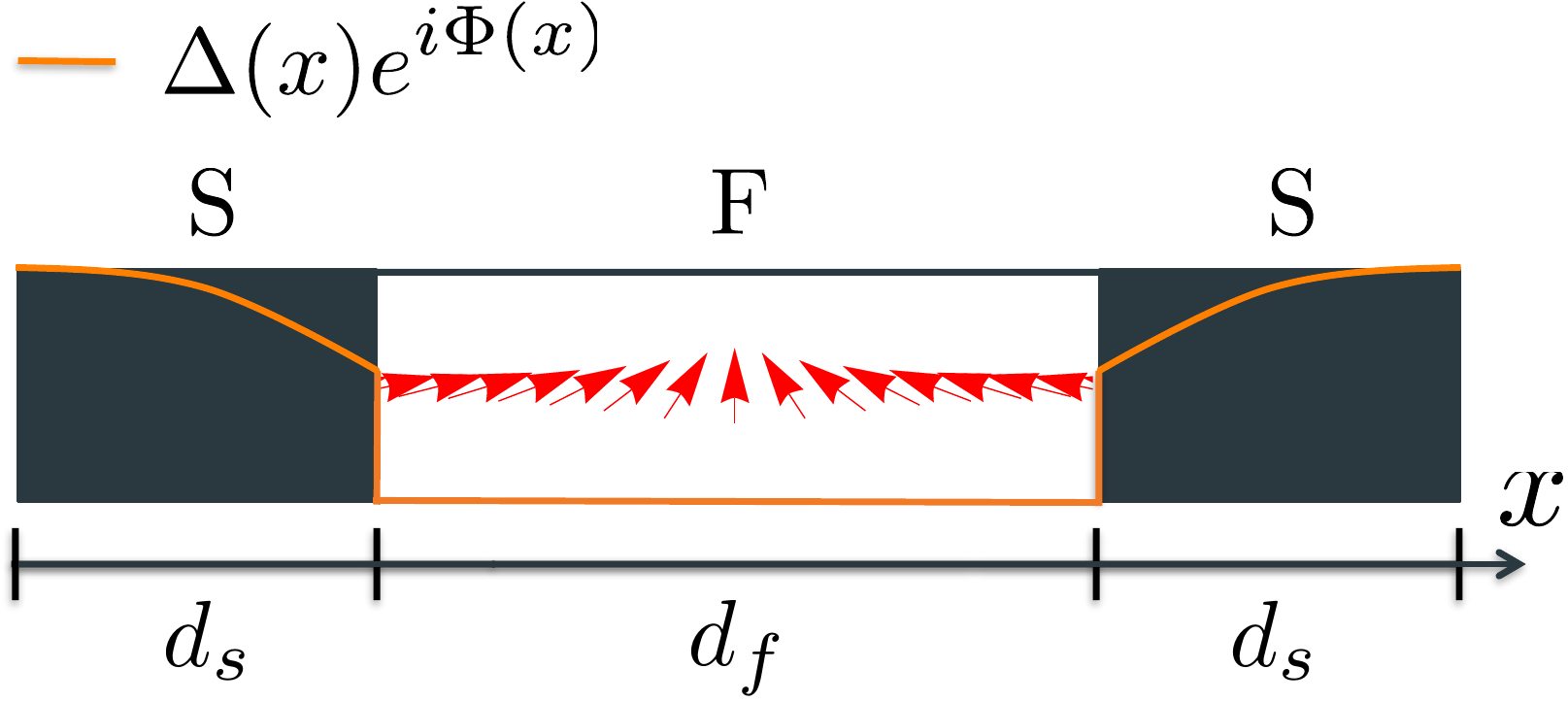}
\caption{Illustration of the SFS structure. The ferromagnetic block exhibits a non-trivial magnetic domain wall structure. The solid line shows a typical variation of the pair-potential within the structure.}
\label{FIG:Illustration of the SFS-Structure}
\end{figure}

We numerically investigate an S-F-S structure that exhibits a magnetic domain wall,  see \Fig{FIG:Illustration of the SFS-Structure}.
Such a structure was treated previously in the Ref.
\onlinecite{Konstandin05}, and later in Ref. \onlinecite{Champel08}. 
We extend the results in Ref. \onlinecite{Konstandin05} by (a) calculating the pair potential self-consistently, and by (b) calculating the current-phase relationships as well as the temperature-dependence of the critical currents. 
%see \Fig{FIG:SFSSNSCriticalCurrents}. 
In addition, we chose a different, normalized, domain wall parameterization, such that the the magnetization vector $\b{J}$ at the start (end) of the ferromagnetic block is always fully polarized in the $+(-)$ $x$-direction.
We keep $|\b{J}|$ constant and spatially vary the orientation, $\b{J}\to\b{J}(x)$, by using the following domain wall parameterization,
\begin{align}
\label{magn0}
J_x&=J \cos\left(\frac{\arctan\left(\frac{x-x_0}{d_w}\right)}{\arctan\left(\frac{d_s-x_0}{d_w}\right)} \frac{\pi}{2}\right),\\
J_z&= J \sin\left(\frac{\arctan\left(\frac{x-x_0}{d_w}\right)}{\arctan\left(\frac{d_s-x_0}{d_w}\right)} \frac{\pi}{2}\right)
\label{magn}
\end{align}
where  $d_w$ is the domain wall width, $J_y=0.0$ and $x_0=d_s+d_f/2$ denotes the middle of the S-F-S structure. The ferromagnetic region extends from $x=d_s$ to $x=d_s+d_f$. \\
We point out that the domain-wall parameterization in Ref. \onlinecite{Konstandin05} had a fixed rotation pitch given by the domain wall thickness, but independent of the thickness of the ferromagnet layer.\cite{corr} Therefore, for higher domain-wall widths the magnetization was already tilted at the interfaces between the ferromagnetic block and the superconductors. Here, we chose a different parameterization by normalizing the argument in the expressions \eqref{magn0}-\eqref{magn} for the magnetization. This is appropriate for the case that the direction of the magnetic moment at the interfaces is determined by magnetic shape anisotropy.\\
The transport equation is given by Eq.\! \eqref{EQ:UsadelEquation} with the replacement $\epsilon \to \left(\epsilon - \b{J}\cdot\b{\sigma}\right)$ to account for the spatial variation of the magnetization. 
In the Riccati parameterization, see Eq.\! \eqref{EqGreenFunction} and Eq.\! \eqref{EQ:Riccatti2}, the equations read:
\begin{align}
&\frac{d^2\gamma}{dx^2}+\left(\frac{d\gamma}{dx}\right) \frac{\mathfraktF}{i\pi}  \left(\frac{d\gamma}{dx}\right)= \nonumber \\
&\frac{i}{D}[\gamma   \DDelta^{*}  \gamma-\l \epsilon\mathit{1}- \b{J}\cdot\b{\sigma}\r   \gamma-\gamma  \l \epsilon \mathit{1}+ \b{J}\cdot\b{\sigma}^{*}\r-\DDelta], 
\\
%\end{align} 
%\begin{align}
&\frac{d^2\tilde{\gamma}}{dx^2}+\left(\frac{d\tilde{\gamma}}{dx}\right)   \frac{\mathfrakF}{-i\pi}  \left(\frac{d\tilde{\gamma}}{dx}\right)= \nonumber \\
&\frac{-i}{D}[\tilde{\gamma}   \DDelta  \tilde{\gamma}+\l\epsilon\mathit{1}+ \b{J}\cdot\b{\sigma}\r  \tilde{\gamma}+\tilde{\gamma}  \l\epsilon \mathit{1}- \b{J}\cdot\b{\sigma}\r-\DDelta^{*}] .
\label{EQ:UsadelSFS}
\end{align} 
At the $S/F$ interfaces $(x_{i}^{S},x_{i}^{F})$ we connect the $\gamma,\tilde{\gamma}$ by continuity conditions: \cite{Konstandin05}
\begin{align}
\gamma(x_{i}^{S})&=\gamma(x_{i}^{F})\\
\frac{d}{dx}\gamma(x_{i}^{S})&=\frac{d}{dx}\gamma(x_{i}^{F}) .
\end{align}
\begin{figure}[t]
\includegraphics[width=8cm]{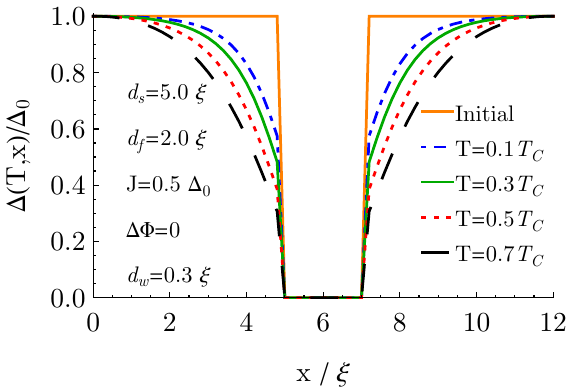}
\caption{Suppression of the pair potential for several temperatures illustrated for an S-F-S structure without phase gradients. 
The solid line (Initial) is the step-like pair potential 
without a self-consistent calculation, depicted for comparison.}
\label{FIG:OrderParameterwithDomainWallStructure}
\end{figure}
\begin{figure}[b]
\includegraphics[width=8cm]{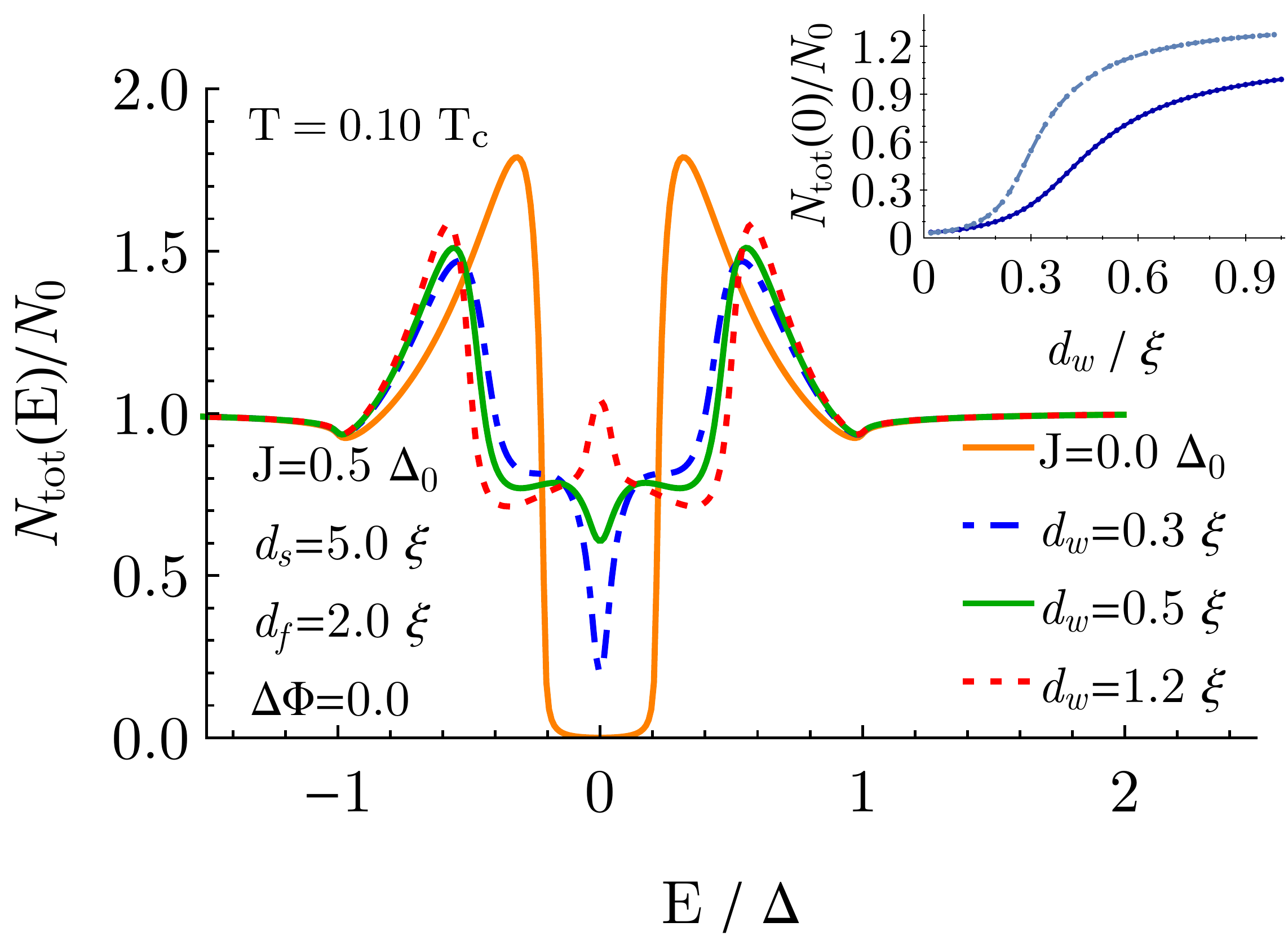}
\caption{Self-consistent LDOS, $N_{\rm tot}$, as a function of energy $E$, normalized to its normal state value $N_0$, for several domain wall widths $d_w$ calculated in the middle of the $F$ region in an S-F-S structure. The case of an S-N-S structure ($J=0$) is shown for comparison as a solid line.
Inset: $N_{\rm tot}$ at the chemical potential ($E=0$) as function of domain wall width $d_w$ for self-consistent pair potential (full line) and for non-self-consistent (step-like) pair potential (dashed lines). }
\label{FIG:DOSAtFermiEnergywPhases}
\end{figure}
The pair potential $\DDelta(T,x)$ is calculated self-consistently according to Eq.\! \eqref{EqSelfConsistencyEquation}. 
This takes into account the suppression of the order parameter close to the ferromagnetic material (inverse proximity effect). In Fig.~\ref{FIG:OrderParameterwithDomainWallStructure} we show the typical behavior of the order parameter as the temperature of the system is varied for a ferromagnet that hosts a domain wall.
%The present calculation is an extension over earlier investigations \cite{Konstandin05} that relied on a step-like form of the potential.
%A strong suppression is observed close to the boundaries of the ferromagnet.

The effect of the domain wall on the local density of states in the system can be seen in \Fig{FIG:DOSAtFermiEnergywPhases}. In comparison to the normal metal $(J=0)$ the minigap is populated with additional Andreev bound states that stem from spin-triplet correlations that are sensitive to the direction of the magnetization.
Non-zero $J_x$ and $J_z$ components convert singlet into triplet amplitudes.\cite{Eschrig11} A non-vanishing $J_x$ induces spin-flips and breaks up a spin-singlet Cooper pair and converts it into an unequal spin-triplet state $\sim\left(\ket{\uparrow\uparrow}-\ket{\downarrow\downarrow} \right)$ whereas $J_z$ induces equal spin-triplet pairings $\sim\left(\ket{\uparrow\downarrow}+\ket{\downarrow \uparrow}\right)$. In the case of increasing domain wall widths, the magnetic domain wall encourages such spin-flip processes and thus creates new Andreev bound states. 
As the domain wall width increases, spectral weight from the shoulders fills up the minigap, as illustrated in Fig.~\ref{FIG:DOSAtFermiEnergywPhases}. 
%As seen there, the minigap is filled progressively with states as the domain wall width increases. 
The inset of Fig.~\ref{FIG:DOSAtFermiEnergywPhases} shows the value of the local density of states at the chemical potential as function of domain wall width.
There is a characteristic value $d_w^\ast $ at which a step-like feature occurs in this plot. 
\begin{figure}[b]
\includegraphics[width=8cm]{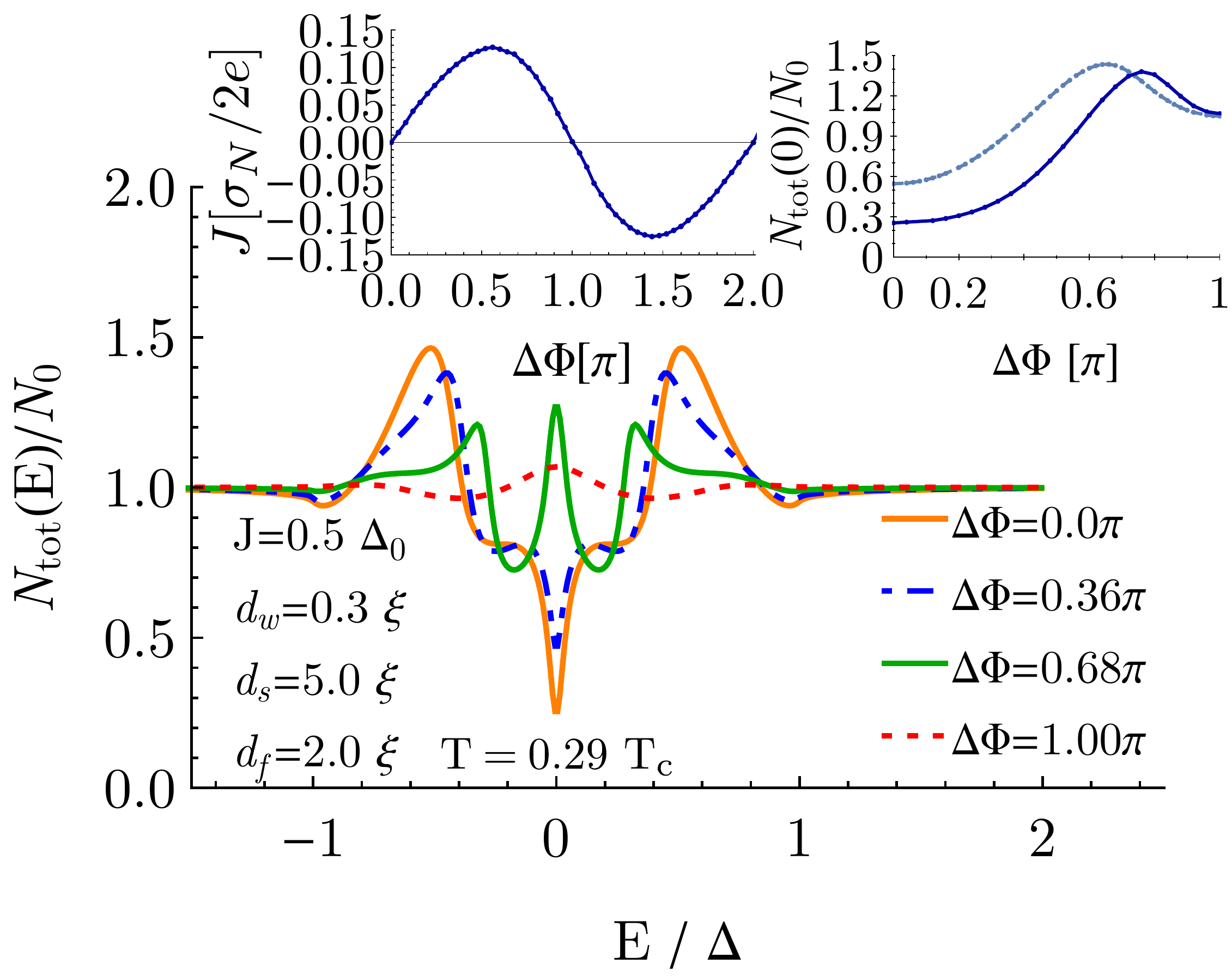}
\caption{Self-consistent LDOS, $N_{\rm tot}(E)$, in an S-F-S structure with domain wall as a function of energy for several phase differences $\Delta \Phi$ between the superconductors, calculated in the middle of the $F$ region, and normalized to the normal state value $N_0$. 
Left inset: current-phase relationship. 
Right inset: $N_{\rm tot}$ at the chemical potential, $E=0$. Full line for self-consistent pair potential. The dashed line shows for comparison the result for a non-self-consistent pair potential.
}
\label{FIG:LDOSvaryDW02phi}
\end{figure}
In comparison to the case of a non-self-consistent pair potential \cite{Konstandin05}, shown as dashed line, this characteristic value $d_w^\ast$ is shifted upwards.
When the magnetic domain wall extends over the whole ferromagnetic region, $J$ varies 
slowly with $x$. This case is similar to the case for a fully polarized ferromagnet. The minigap thus vanishes and local minima appear at  approximately $\pm J/\Delta$. The same effect can be observed for the (S-(F$|$S)-S) structure, where the local density of states is spin-split, see \Fig{FIG:7a}.

For a given domain wall width, we investigate the dependence of the local density of  states on the applied phase gradient, see Fig.~\ref{FIG:LDOSvaryDW02phi}. 
When a finite phase difference is present at the outer elements, supercurrents can flow in the S-F-S-structure. A finite phase difference $\Delta \Phi$ modifies the local density of states as it adds to the phase that is picked up by the quasiparticles during the diffusive motion through the ferromagnet. 
In particular the zero-energy density of states is influenced strongly by the applied phase difference, as it can be seen in the right inset of \Fig{FIG:LDOSvaryDW02phi}. It increases smoothly until a maximum value for $\Delta \Phi <\pi$ is reached. The plot is mirror symmetric around $\Delta \Phi=\pi$ (only values for $\Delta \Phi<\pi $ are shown). For comparison we also reproduce the non-self-consistent result of Ref.~\onlinecite{Konstandin05} as a dashed line in the inset.  We observe that self-consistency of the order parameter gives pronounced corrections to the local density of states, in particular its value at the chemical potential.
Experimentally, tunnel current measurements provide access to the zero-energy density of states.

\begin{figure}[b]
\includegraphics[width=8cm]{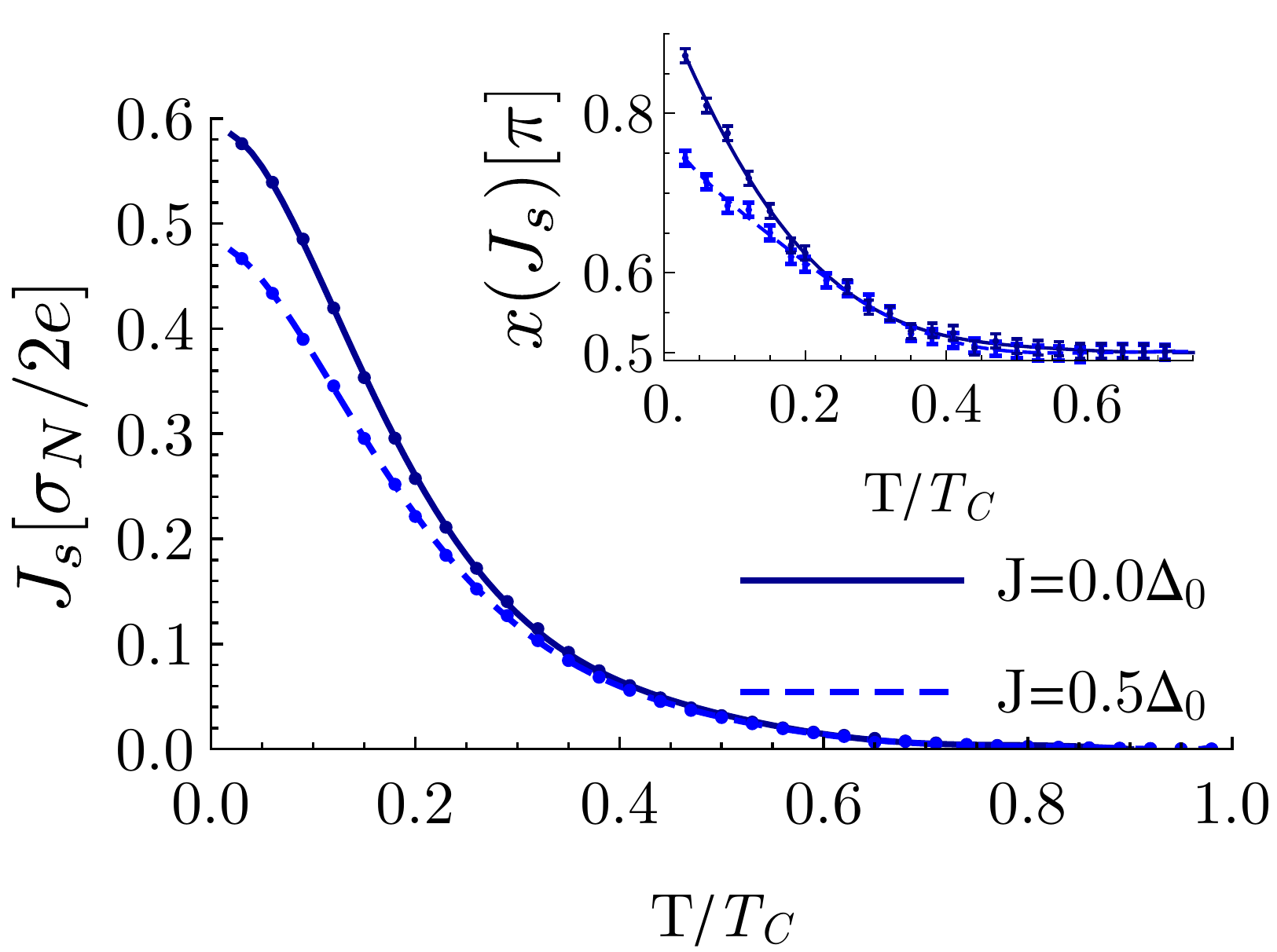}
\caption{Critical Josephson currents for the parameters $d_s=5.0\xi, d_f=2.0\xi$,and a domain-wall of length $d_w=0.3\xi$. The inset shows the phases for which the maximal (critical) current is reached. 
The error bars indicate the numerical uncertainty in determining the extrema in the current-phase relationship. Lines are guides to the eye.}
\label{FIG:SFSSNSCriticalCurrents}
\end{figure}

We also present here self-consistent supercurrents in the S-F-S structure. Supercurrents have not been studied in Ref.~\onlinecite{Konstandin05}.
In Fig.~\ref{FIG:SFSSNSCriticalCurrents}
we plot the temperature-dependence of the critical currents for both a S-N-S structure (full lines) and a S-F-S structure that hosts a domain wall (dashed lines). Additionally, we numerically track the temperature-dependence of the phase difference that leads to the critical current, see inset in \Fig{FIG:SFSSNSCriticalCurrents}.
The critical currents in the S-F-S structure are lowered by a magnetic domain wall in comparison to the case when a magnetic structure is missing, such as is the case in a S-N-S structure. The current-phase relationship in both cases becomes sinusoidal at high temperatures, where the maximum current is reached at a phase difference $\Delta \Phi=\pi/2$. This is reflected in the critical currents as well, as the curves for the S-N-S and the S-F-S structure collapse onto each other approximately when for both cases a sinusoidal current-phase relationship is established. In the low-temperature-regime, the critical currents in the S-N-S structure are offset to higher values than in the case for the S-F-S structure with a domain wall. In both structures however the maximum current is achieved for phase gradients $\Delta \Phi < \pi$, see the inset of \Fig{FIG:SFSSNSCriticalCurrents}. This should be compared to the S-(F|S)-S structure where critical currents are reached for phase differences $\Delta\Phi > \pi$.

%%%%%%%%%%%%%%%%%%%%%%%%%%%%%%%%%%%%%%%%
\section{Conclusion}

Using the model for an S-(S|F)-S Josephson junction depicted in Fig.~\ref{setup},
we have transformed spin-dependent boundary conditions within the (S|F) bilayer into
an effective self energy that enters the Usadel transport equation. 
This allows for a numerically very effective handling of the transport equation. We have used our model to calculate important measurable quantities such as the density of states, spin-magnetizations, the pair potential, and the critical Josephson currents through the system. We also proved that our theory explicitly fulfills the continuity equation, expressing charge conservation, provided self-consistently determined order parameter profiles are used.

We have in particular studied the weak link behavior of such an S-(S|F)-S Josephson junction, showing the characteristic hysteretic current-phase relation \cite{Golubov04}, as indicated by a multi-valued solution. In our case the suppression of superconducting order in the weak-link region is achieved via proximity coupling to a strongly spin-polarized ferromagnet. We study long weak-link structures with a length comparable or larger than the superconducting coherence length. We present a detailed quantitative solution for this problem. We find that self-consistency of the order parameter profile across the weak link is necessary in order to be able to determine the Josephson current in a sensible way.

We also consider a second geometry, an S-F-S junction in which a magnetic domain wall is situated in the center of the F region. 
We have extended previous work\cite{Konstandin05,Bergeret01, Volkov06, Volkov08, Linder09, Alidoust10, Buzdin11, Wu12, Linder14} by
studying in particular the effect of self consistency of the
order parameter in the superconducting leads. We find that self-consistency of the order parameter leads to pronounced modification of the results, in particular the functional dependence of the density of states on domain wall width. We also calculated
the critical Josephson current and find that it is considerably reduced at low temperatures by the presence of a domain wall.

\acknowledgements
J.G. acknowledges financial support by SEPnet/GRADnet during his Euromasters study at Royal Holloway, University of London. J.G. and M.E. appreciate the stimulating atmosphere within the Hubbard Theory Consortium. M.E. acknowledges support by EPSRC (Grant No. EP/J010618/1 and EP/N017242/1).

%%%%%%%%%%%%%%%%%%%%%%%%%%%%%%%%%

%%%%%%%%%%%%%%%%%%%%%%%%%%%%%%%%%%%%%%%%%%%%%%%%%%
% \vspace{-0.5cm}


\begin{thebibliography}{50}
\bibitem{corr}
We note here a typo in Eq. 10 of 
Ref. \onlinecite{Konstandin05}, where it should read
$\theta(z)=-\arctan(4\frac{z-z_0}{d_W})$.
\end{thebibliography}
\end{document}